\documentclass[fleqn,useAMS,usenatbib,usegraphicx]{mn2e}

\usepackage{savesym}   %
\savesymbol{iint}\savesymbol{iiint}
\savesymbol{iiiint}\savesymbol{idotsint}
\usepackage{txfonts}
\restoresymbol{TXF}{iint}
\restoresymbol{TXF}{iiint}
\restoresymbol{TXF}{iiiint}
\restoresymbol{TXF}{idotsint}

\usepackage{upgreek}
\usepackage{amsmath,amssymb}

\usepackage{url}

\defcitealias{goli04}{G04}
\defcitealias{marl07}{M07}
\defcitealias{scvh}{SCvH}

\newcommand{\Si}{S_{\rm i}}
\newcommand{\Li}{L_{\rm i}}  %
\newcommand{\MJ}{{\rm M}_{\rm J}}
\newcommand{\RJ}{{\rm R}_{\rm J}}
\newcommand{\ME}{{\rm M}_{\earth}}

\newcommand{\LSun}{{\rm L}_{\sun}}
\newcommand{\LJ}{{\rm L}_{\rm J}}
\newcommand{\Lbol}{{L_{\rm bol}}}
\newcommand{\LD}{L_{\rm D}}
\newcommand{\Teff}{{T_{\rm eff}}}
\newcommand{\Pphot}{{P_{\rm phot}}}
\newcommand{\Pc}{{P_{\rm c}}}
\newcommand{\rhoc}{{\varrho_{\rm c}}}
\newcommand{\Tc}{{T_{\rm c}}}
\newcommand{\Trcb}{{T_{\rm RCB}}}
\newcommand{\Prcb}{{P_{\rm RCB}}}
\newcommand{\delad}{{\nabla_{{\rm ad}}}}
\newcommand{\tth}{{t_{\rm therm}}}
\newcommand{\kB}{k_{\rm B}}
\newcommand{\EF}{E_{\rm F}}
\newcommand{\Sunits}{\kB/{\rm baryon}}
\newcommand{\Sunitsem}{\kB/{\it baryon}}
\newcommand{\DSsyst}{\Delta S_{\rm syst}}

\usepackage{xcolor}
\newcommand{\Ae}[1]{#1}

\begin{document}

\title[Constraining initial entropies of exoplanets] {  %
       Constraining the initial entropy of directly-detected exoplanets
       }
\author[G.-D.~Marleau and A.~Cumming] {
        G.-D.~Marleau$^{1,2}$\thanks{E-mail: marleau@mpia.de, cumming@physics.mcgill.ca} and
        A.~Cumming$^2$\footnotemark[1] \\
        $^1$Max-Planck-Institut f\"ur Astronomie, K\"onigstuhl 17, D-69117 Heidelberg, Germany\\
        $^2$Department of Physics, McGill University, 3600 rue University, Montr\'eal, Qu\'ebec H3A 2T8, Canada
        }

\maketitle

\begin{abstract}
The \Ae{post-formation,} initial entropy $\Si$ of a gas giant planet is a key witness to its mass-assembly history and
a crucial quantity for its early evolution. However, formation models are not yet able to predict reliably $\Si$,
\Ae{making unjustified the use solely of traditional, `hot-start' cooling tracks to interpret direct imaging results and calling for}
an observational determination of initial entropies to guide formation scenarios.
Using a grid of models in mass and entropy, we show how to place
joint constraints on the mass and initial entropy of an object from its observed luminosity and age.
\Ae{This generalises the usual estimate of only a lower bound on the real mass    through hot-start tracks.}
Moreover, we demonstrate that with mass
information, e.g.\ from dynamical stability analyses or radial velocity, tighter bounds can be set on the initial entropy.
We apply this procedure to 2M1207~b and find that its initial entropy is at least 9.2~$\Sunits$, assuming that it does not burn deuterium.
For the planets of the HR~8799 system, we infer that they must have formed with $\Si > 9.2~\Sunits$, independent of uncertainties about the age of the star.
Finally, a similar analysis for $\beta$~Pic~b reveals that it must have formed with $\Si >10.5~\Sunits$, using the radial-velocity mass upper limit.
These initial entropy values are respectively ca.\ 0.7, 0.5, and 1.5~$\Sunits$ higher than the ones obtained from core accretion models
by Marley et al.,    %
thereby {\em quantitatively} ruling out the coldest starts for these objects and constraining warm starts, especially for $\beta$~Pic~b.
\end{abstract}


\section{Introduction}

While only a handful of directly-detected exoplanets is currently known, the near future should bring a
statistically significant sample of directly-imaged exoplanets, thanks to a number of surveys underway or coming
online soon. Examples include VLT/SPHERE, Gemini/GPI, Subaru/HiCIAO, Hale/P1640 %
\citep{vigan10,chauv10,mcb11,mcelw12,yama13,hink11_pasp}.   %
An important input for predicting the yields of such surveys and for interpreting their results is the cooling
history of gas giant planets.
In the traditional approach \citep{stevenson82,burr97,bara03},
objects begin their cooling, fully formed, with an arbitrarily high specific
entropy\footnote{\Ae{The `initial' entropy refers to the entropy at the beginning of the pure cooling phase,
once the planet's mass is assembled.}}
and hence radius and luminosity. 
In the past, the precise choice of initial conditions \Ae{for the cooling} has been of no practical consequence because only the 4.5-Gyr-old
Solar System's gas giants were known, while high initial entrop\Ae{ies 
are} forgotten on the short Kelvin--Helmholtz time-scale $GM^2/RL$   %
\citep{stevenson82,marl07}. %
However, direct-detection surveys are aiming specifically at young ($\leqslant500$~Myr) systems, and the traditional models,
as their authors explicitly recognised, are not reliable at early ages.

Using the standard core-accretion formation model \citep{pollack96,boden00,hubic05},
\citet[hereafter \citetalias{marl07}]{marl07} %
found that newly-formed gas giants produced by core accretion should be substantially colder than \Ae{what} the usual cooling
tracks that begin with arbitrarily hot initial conditions assume.
These outcomes are known as `cold start' and `hot start', respectively. 
\citetalias{marl07} however noted that there are a number of assumptions and approximations that go into the
core accretion models that make the predicted entropy uncertain. In particular, the accretion shock at the surface
of the planet is suspected to play a key role as most of the mass is processed through it
(\citetalias{marl07}; \citealp*{bara10}) but there does
not yet exist a satisfactory treatment of it.
Furthermore, there may also be an accretion shock in the other main formation scenario, gravitational instability,
such that it too could yield planets cooler than usually expected \citep[see section~8 of][]{morda12_I}.
\Ae{Conversely, \citet{morda13} recently found that the initial entropy varies strongly with the mass
of the core, leading nearly to warm starts for reasonable values of core masses in the framework of core accretion.}

The most reasonable viewpoint for now is therefore to consider that the initial entropy is highly uncertain and may lie
almost anywhere between the cold values of \citetalias{marl07} and the hot starts. In fact, \citetalias{marl07} calculated
`warm start' models that were intermediate between the cold and hot starts, and recently, \citet{spiegel12}
calculated cooling tracks and spectra of giant planets beginning with a range of initial entropies.
This uncertainty in the initial entropy means that observations are in a privileged position to %
inform models of planet formation, for which the entropy (or luminosity) \Ae{at the end of the accretion phase} is an easily accessible output.
Its determination both on an individual basis as well as for a statistically meaningful sample of planets would be very valuable,
with the latter allowing quantitative comparisons to %
planet population synthesis \citep{ida04,morda12_I,morda12_II}.
\Ae{Moreover, upcoming and ongoing surveys should bring, in a near future, both qualitative and quantitative changes
to the collection of observed planets to reveal a number of light objects at moderate separations from their parent star
(i.e.\ Jupiter-like planets), which contrasts with the few currently known direct detections.}
\emph{\Ae{With this in mind, we stress} the importance of interpreting direct observations in a model-independent fashion}
and of thinking about the information that these yield about the initial (i.e.\ post-formation) conditions of gas giants.

In this paper, we investigate the constraints on the mass $M$ and initial, \Ae{post-formation} entropy $\Si$ that come from a luminosity and age
determination for a directly-detected object, focusing on exoplanets.
Since the current luminosity increases \Ae{monotonically} both with mass and initial entropy,
these constraints take the form of a `trade-off curve' between $M$ and $\Si$.
We show that, in the planetary regime, the allowed values of $M$ and $\Si$ can be divided into two regions.
The first is a hot-start region where the initial entropy can be arbitrarily high but where the mass is \Ae{essentially unique}.
This corresponds to the usual mass determination by fitting hot-start cooling curves.
The second corresponds to solutions with a lower entropy \Ae{in a narrow range} and for which the planet mass has to be
larger than the hot-start mass.
\Ae{(A priori, this may even reach into the mass regime where deuterium burning is important,
which is discussed in a forthcoming paper.)}   %
The degeneracy between mass and initial entropy means that in general the mass and entropy cannot be constrained independently
from a measurement of luminosity alone. When additional mass information is available, for example for multiple-planet systems
or systems with radial velocity, it is possible to constrain the formation entropy more tightly.

We start in Section~\ref{sec:struct+cool} by describing our gas-giant cooling models,
discussing the luminosity scalings with mass and entropy, and comparing to previous calculations in the literature.
In Section~\ref{sec:genconstr}, we show in general how to derive constraints on the mass and initial entropy of a directly-detected exoplanet
by comparing its measured luminosity and age to cooling curves with a range of initial entropies.
We briefly consider in Section~\ref{sec:gTconstr} similar constraints based on a (spectral) determination of the effective temperature
and surface gravity.
\Ae{After a brief discussion of the luminosity--age diagram of directly-detected objects},
we apply the procedure based on bolometric luminosities in Section~\ref{sec:obsobj}
to three particularly interesting systems -- 2M1207, HR~8799, and $\beta$~Pictoris --
and derive lower bounds on the initial entropies \Ae{of the companions}. For HR~8799 and $\beta$~Pic, we use the available information on the masses
to constrain the initial entropies more tightly. Finally, we offer a summary and concluding remarks in Section~\ref{sec:discussion}.

\section{Cooling models with arbitrary initial entropy}
\label{sec:struct+cool}

Given its importance in the initial stages of a planet's cooling, we focus in this paper on the internal entropy of
a gas giant as a fundamental parameter which, along with its mass, determines its structure and controls its evolution.
Very few authors have explicitly
considered entropy as the second fundamental quantity even though this approach is more transparent than the usual,
more intuitive use of time as the second independent variable.
In this section, before comparing our models to standard ones, we discuss interior temperature--pressure profiles
and semi-analytically explain the different scalings of luminosity on mass and entropy %
which appear in the models.

\subsection{Calculation of time evolution} %
\label{sec:coolingmodels}
We calculate the evolution of cooling gas giants by a `following the adiabats' approach \citep{hubb77,forthubb04,ab06}.
We generate a large grid of planet models with ranges of interior specific entropy $S$ and mass $M$
and determine for each model the luminosity $L=L(M,S)$ at the top of the convection zone.
We then use it to calculate the rate of change of the convective zone's entropy ${\rm d}S/{\rm d}t$, which is defined by
writing the entropy equation $\partial L/\partial m = -T\partial S/\partial t$ as
\begin{equation}\label{eq:L}
 L = -\frac{{\rm d}S}{{\rm d} t}\int\! T\, {\rm d}m + \LD,
\end{equation}
with $\LD$ the luminosity due to deuterium burning in the convection zone.
\Ae{With the current $S$ and ${\rm d}S/{\rm d}t$ in hand, calculating a cooling curve
becomes a simple matter of stepping through the grid of models at a fixed mass.
This is computationally expeditious and gives results nearly identical to the usual procedure
based on the energy equation \citep{henyey64,kipp}, as discussed in Section~\ref{sec:modelcomp}.}

The assumptions in \Ae{the `following the adiabats'} approach are that $\partial S/\partial t$ is constant throughout the convective zone
and that no luminosity is generated in the atmosphere.
For these to hold, we require both the thermal time-scale $\tth(r) = Pc_PT/gF$ \citep{ab06} --
where $c_P$ is the specific heat capacity, $g$ the gravity, and $F$ the local flux -- in the radiative zone
overlying the convective core and the convective turnover\footnote{We do make the standard assumption that the interior
is fully convective, even though stabilising compositional gradients
have been suggested to possibly shut off large-scale convective motions \citep{stevenson79,leconte12,leconte13}.}
time-scale to be much shorter than the time-scale on which the entropy is changing,
$\tau_S=-S/({\rm d}S/{\rm d}t)=M\overline{T}S/L$, with $\overline{T}$ the mass-weighted mean temperature in the convection zone.
A similar approach was used to study the evolution of ohmically-heated irradiated gas giants by \citet{huang12}.

We do not include deuterium burning directly in the planet models since it always occurs inside the convection zone.
Instead, we calculate the deuterium burning luminosity per unit deuterium mass fraction for each model in
the grid and use it in equation (\ref{eq:L}) to follow the cooling of the planet and the time evolution of the
deuterium mass fraction averaged over the convective region.
\Ae{However, we focus in this work on masses below the (parameter-dependent) deuterium-burning limit near $12$--14~$\MJ$
(\citealp*{spiegburrmils11}; \citealp{moll12,boden13}),
such that deuterium burning does not play a role in the objects' evolution.}
We defer a detailed exploration of deuterium burning in our models to an ulterior publication  %
but already mention \Ae{the interesting result} that cooling curves starting at low entropy can exhibit an initial increase
in their luminosity due to deuterium burning.
We subsequently noticed that the colder starts in fig.~8 of \citet{moll12} show a similar behaviour,
\Ae{and that \citet{boden13} also find in their formation simulations the possibility of deuterium `flashes'.}

\subsection{Calculation of gas giant models}
\label{sec:MSmodels}
To construct a model with a given mass $M$ and internal entropy $S$, we integrate inwards from the photosphere and outwards
from the centre, adjusting the central pressure, the cooling time $\tau_S$, and the luminosity $L$ and the radius $R$
at the top of the inner zone until the two
integrations match at a pressure of 30~kbar. We use the Eddington approximation at the photosphere, setting $T=\Teff$
at $\Pphot=2g/3\kappa$, and take the solar-metallicity \citep[based on][]{lodders03}
radiative opacity from \citet*{freed08}. In the deep interior,
the contribution from the electron conductive opacity of \citet{cassisi07} is also included.
The equation of state for the hydrogen--helium mixture is that of \citet*[hereafter \citetalias{scvh}]{scvh} with
a helium mass fraction $Y=0.25$. Since we focus on gas giants, we do not include a rocky core, which was found in a test grid to
increase the luminosities by at most a few per cent at the lower masses, as in \citet{saumon96}.

The grid %
has a lower entropy limit of $S\simeq7$--$7.5~\Sunits$\footnote{Entropy values in this work are given in the
usual units, \Ae{written explicitly or not,} of multiples of Boltzmann's constant $\kB$ per baryon (i.e.\ per mass of hydrogen atom).
For comparison, 4.5-Gyr-old Jupiter has a current entropy of $7~\Sunits$ \Ae{(\citealp{marl07}, but see Appendix~\ref{sec:Sdiff}).}}.
The upper $S$ limit in the grid is set by the requirement that $R \la 7~\RJ$, a value found to make numerical
convergence straightforward. The upper limits are $S_{\rm max}\simeq 12$ near 1~$\MJ$ and $S_{\rm max}\simeq 14$
for the higher masses.

\begin{figure}
\includegraphics[width=84mm]{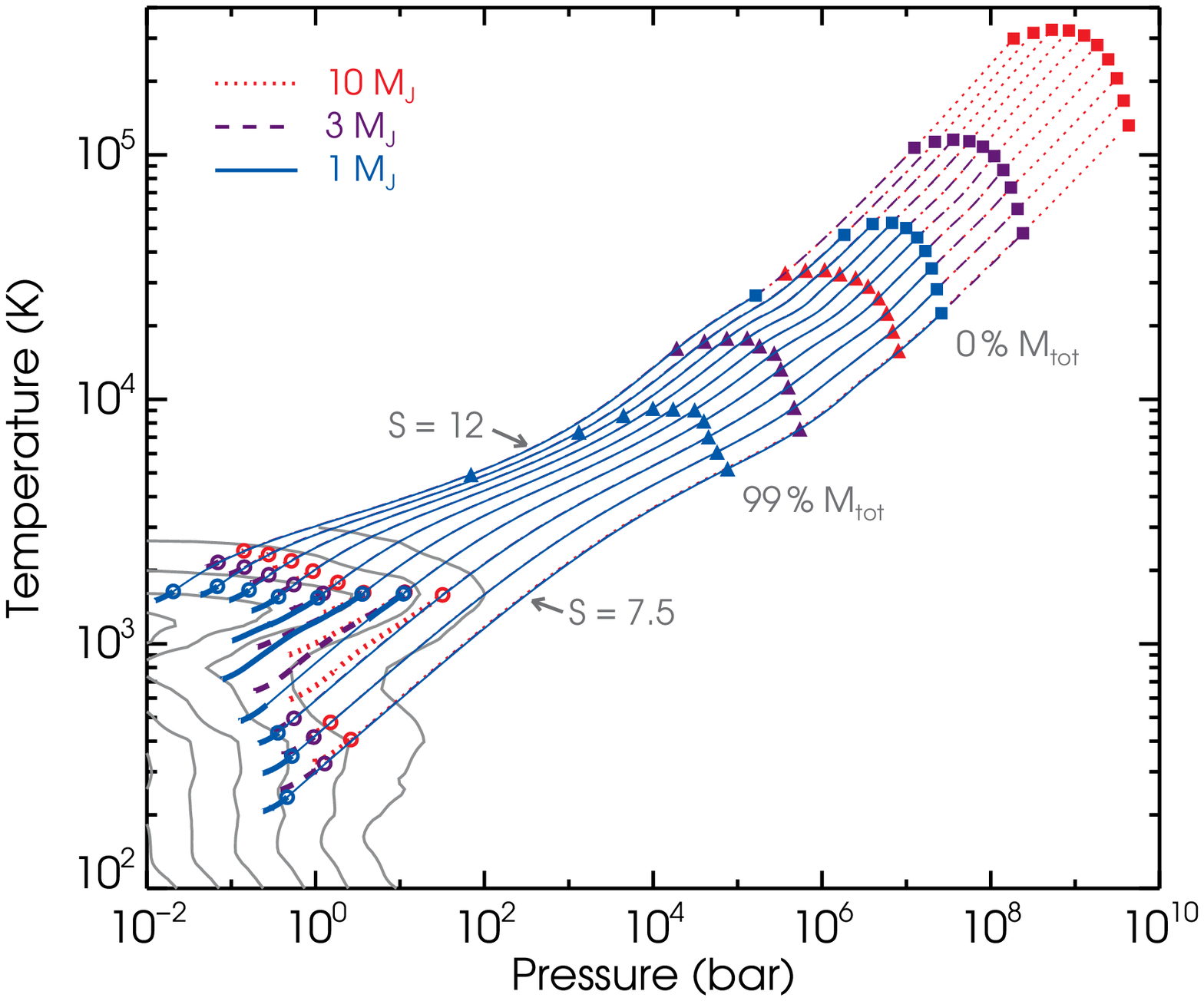}\\
\includegraphics[width=84mm]{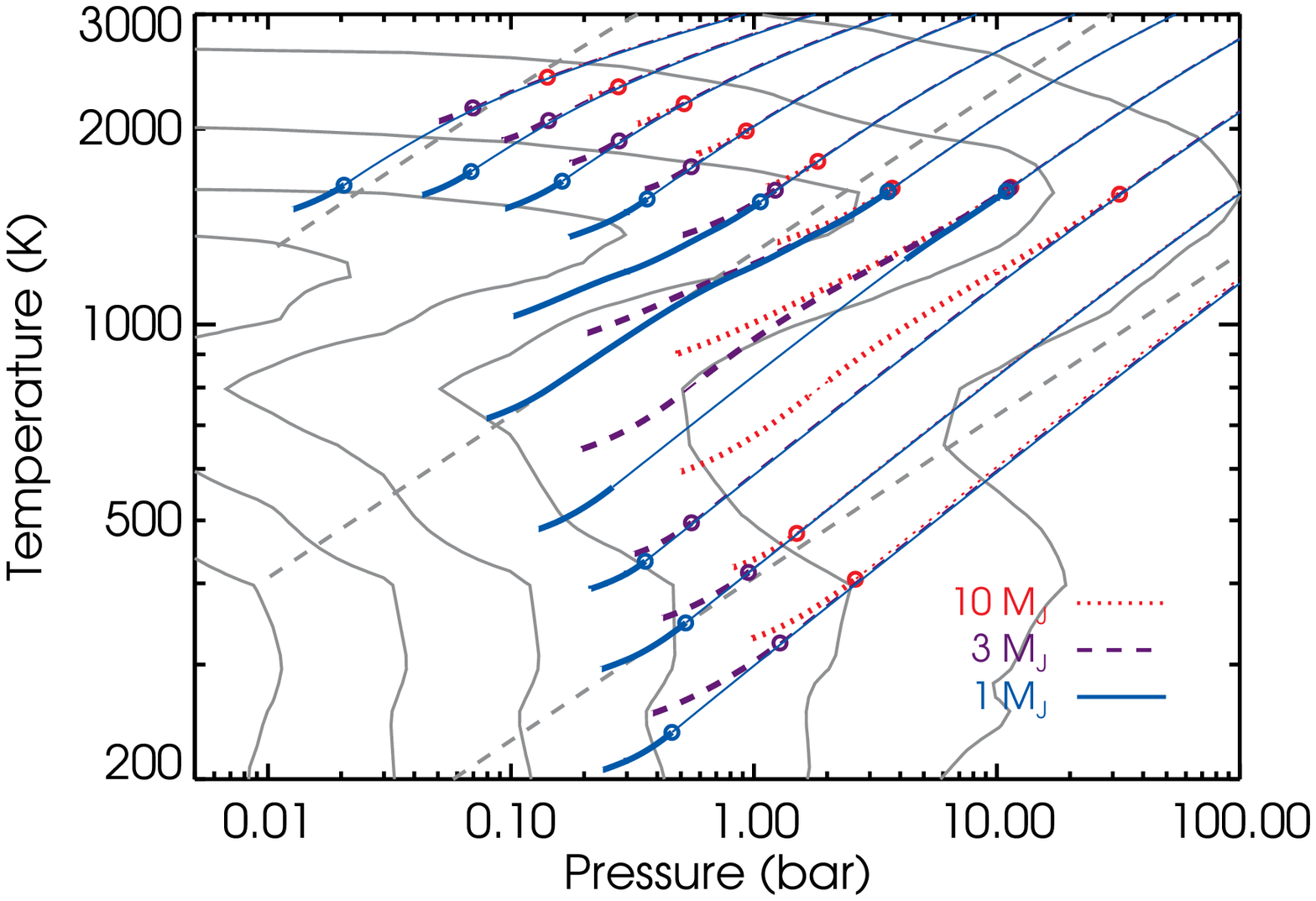}
\caption{
{\em Top panel:}
Temperature--pressure profiles of core-less planets with entropy $S=7.5$--$12~\Sunits$ (bottom to top), in steps of 0.5~$\Sunits$,
and $M=1$, 3, and $10~\MJ$ (full, dashed, and dotted lines, respectively).
Squares indicate the centre and triangles mark where the enclosed mass is 99 per cent of the total mass for each model.
Curves end at the photosphere in the Eddington approximation, and
the thick part(s) of each curve show the radiative zone(s), while the thin part(s) of the curve are convective;
the radiative-convective boundary relevant for the thermal evolution is highlighted by a ring.
Thin grey lines at low $P$ and $T$ indicate contours of constant \Ae{Rosseland mean} opacity \citep{freed08},  %
with $\log \kappa = -3.5$ to $-1.0$~\mbox{(cm$^2$\,g$^{-1}$)} in steps of 0.5~dex, increasing with $P$.
\Ae{See text for a mention of the dominant sources of opacity in some parts of the diagram.}
\Ae{
  {\em Bottom panel:} Same as top panel but focussing on the atmospheres. The three diagonal dashed lines
  mark $\log R \equiv \log \varrho/{T_6}^3 = 2$, 4, and 6~\mbox{(g\,cm$^{-3}$\,K$^{-3}$)}, from top to bottom. %
}
}
\label{fig:jupcomp}
\end{figure}

Fig.~\ref{fig:jupcomp} shows interior profiles in the $T$--$P$ plane for a range of entropies and masses.
Schematically, since hydrostatic balance dictates $\Pc\sim M^2/R^4$, increasing the mass at a fixed entropy
extends the centre to a higher pressure along the adiabat. This is exacerbated at high entropies,
where the planet substantially shrinks with increasing mass, while low-entropy objects have a roughly constant radius.
\Ae{(Radii as a function of $M$ and $S$ are presented in Appendix~\ref{app:R}.)}
At fixed mass, increasing the entropy mostly shifts the centre to higher $T$ or to lower $P$. The first case obtains
for low-entropy objects, which are essentially at zero temperature in the sense that $\kB \Tc \ll \EF$,
where $\Tc$ is the central temperature and $\EF$ is the Fermi energy level at the centre,
\Ae{taken to approximate the electron chemical potential}.
Increasing the entropy partially lifts the degeneracy since the degeneracy parameter
$y \approx \kB \Tc/\EF \propto \Tc/{\rhoc}^{2/3}$ is a monotonic function of $S$,
and $\Pc$ remains constant because of the constant radius.
At entropies higher than a turn-over value of $\simeq10.4~\Sunits$, the central temperature does not increase
(and even decreases) with entropy. %
\Ae{As pointed out in \citet{paxton13}, this entropy value is given by $\kB \Tc\sim \EF$ -- we find that $y\simeq0.15$
for $S=10.4$ -- and is thus independent of mass.}
As for $\Pc$, it decreases because the radius increases.
These behaviours also hold at higher masses not shown in the figure.

As for the 1-$\MJ$ planet with $S=9$ shown in Fig.~\ref{fig:jupcomp}, some models with entropy $S\simeq8$--9.5 show a second,
`detached' convective zone at lower pressures, which follows from a re-increase of the Rosseland mean opacity
(see section~3.1 of \citealp{burr97}).
This second convection zone, which is at most at an entropy 0.2~$\Sunits$ higher than the convective core,
will not affect the evolution of the object
since the radiative thermal time-scale is much shorter than the cooling time $\tau_S$ throughout the atmosphere.  %
This holds in particular at the inner radiative-convective boundary (RCB);
for example, the 1-$\MJ$, $S=10$ model has $\tau_S = 14$~Myr and $\tth = 0.2$~Myr at its RCB.
\Ae{Planets with a second convective zone can equivalently be thought of as having a radiative shell interrupting their convective zone,
as originally predicted for Jupiter's adiabat by \citet{guillot94b,guillot94a} based on too low opacities
(see e.g.\ the brief review in \citealp{freed08}).}

Finally, we note that the higher-entropy objects ($S\ga11.5$, with some dependence on mass)
are convective from the centre all the way to the photosphere.  %

\subsection{Luminosity as a function of mass and entropy}
\label{sec:L(M,S)}
\begin{figure}
\includegraphics[width=84mm]{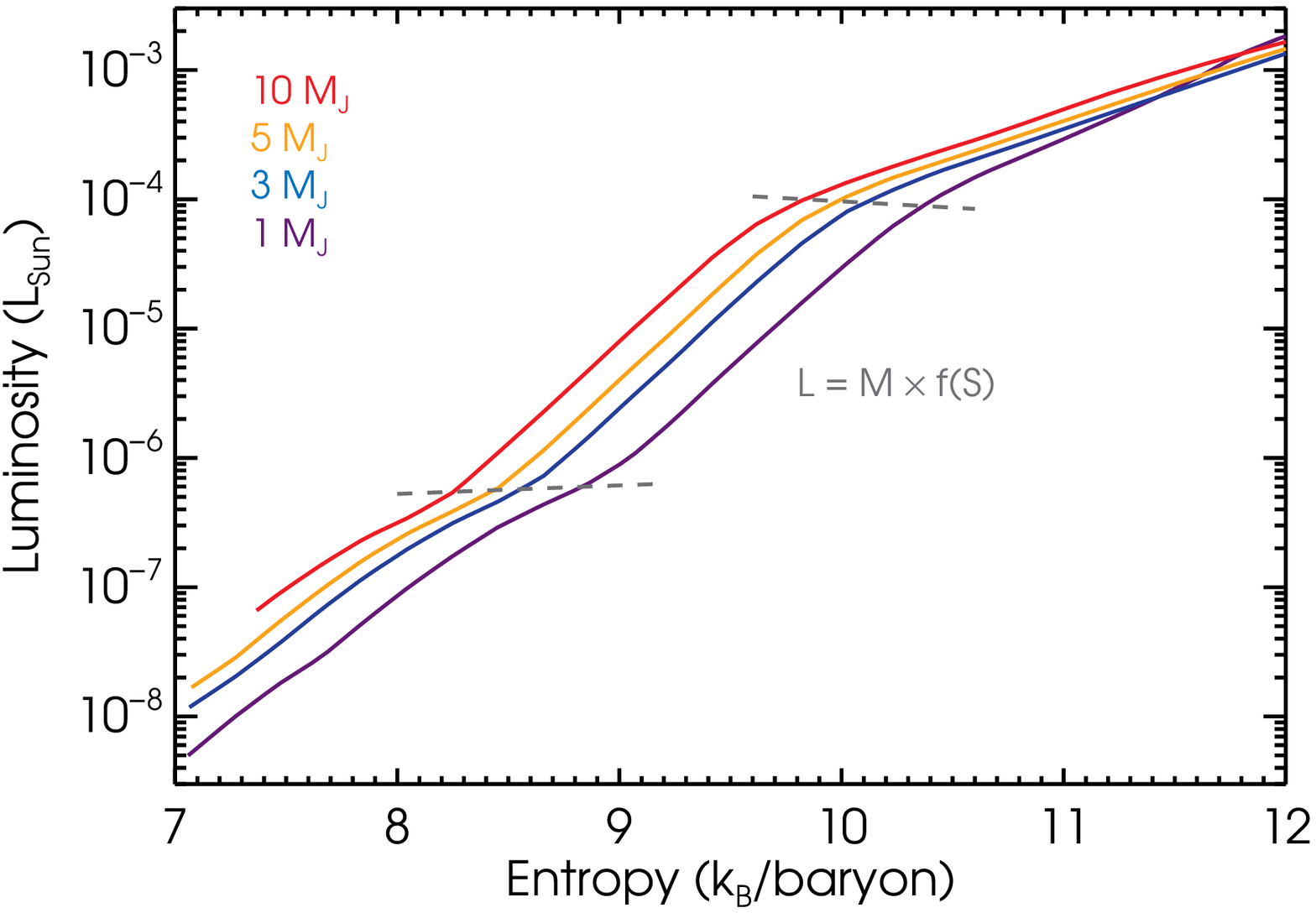}
\caption{
Luminosity as a function of entropy for our planet models. The masses shown are (bottom to top)
$M=1,3,5,$ and 10~$\MJ$.\Ae{The two dashed lines approximately indicate the mass-dependent entropy
boundaries between the low-, intermediate- and high-entropy regimes of the luminosity $L=L(M,S)$.
At intermediate entropies, the atmosphere reaching the radiative-zero solution
leads to the simple result that $L/M$ depends only on $S$ (see text for details).}
Here as throughout, $L_{\rm Sun} = 3.86\times10^{33}$~\mbox{erg\,s$^{-1}$}.
}
\label{fig:SL}
\end{figure}

To provide a model-independent way of thinking about an object's brightness
and thus to facilitate comparison with other models,   %
we show in Fig.~\ref{fig:SL} the luminosity of the planet as a function
of its internal entropy. We focus on objects without significant deuterium burning. Two main regimes are apparent: %
at lower entropies, the scaling with mass is roughly $L\sim M$,
while at high entropies, the luminosity becomes almost insensitive to mass. %
Looking more closely, the high-luminosity regime is described by $L\propto M^{0.3}$ (for $M\ga1~\MJ$), and
the brief \Ae{steepening} of the luminosity slope with \Ae{respect to} entropy \Ae{between $L\simeq10^{-6}$ and $10^{-4}~\LSun$}
(at $S\sim8.2$--9, depending on mass) marks a transition from $L\propto M^1$ to $L\propto M^{0.7}$
at fixed intermediate and low entropy \Ae{respectively}.

To try to understand these luminosity scalings,
we firstly note that the radiative luminosity at the radiative-convective boundary,
which is equal to the total luminosity, can always be written as \citep{ab06}
\begin{equation}
\label{eq:Lrcb}
 L=\left(4\upi r^2 \frac{4acT^3}{3\kappa\rho} \frac{{\rm d}T}{{\rm d}r}\right)_{\rm RCB}= 
     \frac{16\upi Gac}{3} \frac{MT_{\rm RCB}^4 \delad_{\rm RCB} }{ \kappa_{\rm RCB} \Prcb},
\end{equation}
approximating the convection zone to contain the whole mass (cf.\ \Ae{the 99-per-cent mass} labels in Fig.~\ref{fig:jupcomp}) and radius.
Thus, one way of obtaining $L(M,S)$ is to express the four quantities $\Prcb$, $\Trcb$,
$\delad_{\rm RCB}$ and $\kappa_{\rm RCB}$ in terms of $M$ and $S$.
This will now be done for the three regimes in turn, starting with low entropies.

\subsubsection[Low-entropy regime]{Low-entropy regime: \ensuremath{L\propto M^{0.7}f_{\rm low}(S)}}
\label{sec:Lscalinglow}
To begin, consider the entropy dependence of luminosity at a fixed mass when $S\la9$--8 (at 1--10~$\MJ$ respectively).
Fig.~\ref{fig:jupcomp} reveals that the opacity at the radiative-convective boundary $\kappa_{\rm RCB}$ remains constant at a given mass,
which provides a first relation. Secondly, over the temperature and pressure range
of interest in this regime, the hydrogen and helium remain respectively molecular and neutral. %
This implies that $\delad_{\rm RCB}$ is constant and that the entropy has a simple functional form in the ideal-gas approximation,
given by the Sackur--Tetrode expression \citep[e.g.][]{callen85}. For an H$_2$--He~{\sc i} mixture
with $Y=\frac{1}{4}$, this is  %
\begin{equation}
\label{eq:SackurTetrode}
S = 9.6 + \frac{45}{32}\ln (T/1600~{\rm K}) -\frac{7}{16}\ln (P/3~{\rm bar}) %
\end{equation}
(hence $\delad=0.31$), where \Ae{$S$ is the entropy per baryon in multiples of $\kB$},
$\ln 10=2.3$, and the reference $T$ and $P$ values were chosen for Section~\ref{sec:Lscalingmid}.
Finally, $\Trcb$ is approximately fitted by $\Trcb \propto 10^{0.19S}$.
Combining the four relations (constant $\delad_{\rm RCB}$ and the three non-trivial ones)
with equation (\ref{eq:Lrcb}) yields $L \propto 10^{1.04S}$.
This is quite close to a power-law fit of Fig.~\ref{fig:SL}, which gives $L \propto 10^{1.3S}$.

Next, consider the position of the RCB at fixed low entropy for different masses. Fixed entropy
immediately implies $T \propto P^{\delad}$, and the constant $\delad=0.31$ gives the second relation.
The profiles of Fig.~\ref{fig:jupcomp} indicate that in this low-entropy regime \Ae{(for $T\simeq200$--400~K,
$P\simeq0.01$--10~bar)}, the opacity depends
approximately only on the pressure, with $\kappa\propto P^{0.9}$. Finally, we find that $\Prcb\propto M^{0.5}$, %
thus providing the fourth relation.
Combining these and using equation (\ref{eq:Lrcb}) gives $L\propto M^{0.8}$, in good agreement with a direct fit,
which gives $L\propto M^{0.7}$.

Thus, a constant adiabatic gradient, an RCB opacity dependent only on the mass, and a few power laws
suffice to show that in the low-entropy regime
\begin{equation}
 L_{\rm low} = 10^{-7.7}~\LSun \, \left(\frac{M}{\MJ}\right)^{0.7}\,10^{1.3(S-7.5)}, 
\end{equation}
which defines $f_{\rm low}(S)$ up to a constant. For Jupiter's adiabat with $T=165$--170~K at 1~bar \citep{saumon04}
and thus an entropy of 6.71--6.75~$\Sunits$,   %
this predicts $\simeq1.9\times10^{-9}~\LSun$, 
in good agreement with the current value $\LJ =8.7\times10^{-10}~\LSun$.

\subsubsection[Intermediate-entropy regime]{Intermediate-entropy regime: \ensuremath{L\propto Mf_{\rm rz}(S)}}
\label{sec:Lscalingmid}
We now look at intermediate entropies, which are in the range 9--10.2 at 1~$\MJ$ to 8.2--\Ae{9.6}~$\Sunits$ at 10~$\MJ$.
\Ae{(More concisely, this corresponds to planets with $L\sim 10^{-6}$--$10^{-4}~\LSun$.)}
As explained by \citet{ab06}, equation (\ref{eq:Lrcb}) is
of the form $L=M f(S)$, where $f(S)$ is a function of the entropy in the convection zone, only if the quantity
$T^4\delad/\kappa P$ at the RCB is a function of a unique variable, $S$.   %
The intermediate-$S$ behaviour can then be understood by noticing in Fig.~\ref{fig:jupcomp} that
objects at those entropies have a $T_{\rm RCB}$ independent of internal entropy
\emph{and} an extended atmosphere (interrupted or not by a second convection zone, \Ae{i.e.\ including the deeper radiative window when present}).
For these planets, the photosphere is sufficiently far from the RCB ($\Prcb \ga 10\,\Pphot$, $\Trcb^4 \ga 5\,\Teff^4$)
that the atmosphere merges on to the radiative-zero profile, which solves \citep{cox68} %
\begin{equation}
\label{eq:rz}
\frac{{\rm d}T}{{\rm d}P} = \frac{3}{16acG}\frac{L}{M}\frac{\kappa}{T^3},
\end{equation}
where the boundary condition is by definition of no consequence.
The solution is a $T(P)$ relation, which yields the radiative gradient $\nabla_{\rm rad} = {\rm d}\ln T/{\rm d}\ln P$ along it.
The important point is that both the atmosphere profile and its gradient depend only on the quantity $L/M$.
Now, choosing an internal entropy fixes the adiabat, i.e.\ sets a second $T(P)$ relation.
We require that at the intersection of the atmosphere and the adiabat
$\nabla_{\rm rad}$ be equal to $\nabla_{\rm ad}$, which is the slope of the chosen adiabat.
This thus pins down $\nabla_{\rm rad}$ in the atmosphere and hence $L/M$ (and the atmosphere profile itself).
Therefore, there is a unique $L/M$ associated with an $S$, which means that $L$ must be of the form\footnote{This
regime was obtained by \citet{ab06} when looking at irradiated planets.
This can be roughly thought to fix $\Trcb$ to the irradiation temperature,
so that $T^4\delad/\kappa P$ at the RCB is automatically a function of only one thermodynamic variable,
for instance $S$, for all entropies.
} $L/M=f(S)$.

Since we \Ae{determine convective instability through the} Schwarzschild criterion $\delad < \nabla_{\rm rad}$,
where $\delad$ is relatively constant and the radiative gradient is given by
\begin{equation}
\label{eq:delrad}
 \nabla_{\rm rad} = \frac{3LP}{16\upi acGMT^4}\kappa
\end{equation}
near the surface, a slow inward increase of $\kappa$ will ensure a deep RCB.
Consequently, the $L\propto M f(S)$ scaling will hold for those $M$ and $S$
such that, starting at the photosphere with $\Pphot\simeq f(M)$ and $\Teff \simeq f(\Pphot,S)$,
the opacity increases only slowly along the adiabats. (For a power-law opacity $\kappa=\kappa_0P^nT^{-n-s}$, this means $n/(n+s)\simeq\delad$.)
This is indeed the case for intermediate-$S$ models in Fig.~\ref{fig:jupcomp}, where the $T(P)$ profiles are nearly aligned
with contour lines of constant opacity.
For this argument to hold,
the radius must be rather independent of the mass \citep[cf.][]{zapol69} and of the entropy.
Also, one needs to approximate the entropy at the photosphere to be that of the interior,
which is reasonable: even in the extended atmospheres, the entropy increases by at most $\simeq 0.5~\Sunits$
\Ae{over our grid of models}.

The functional form of $f(S)$ can be obtained by fixing $\Trcb$ and using equation (\ref{eq:SackurTetrode})
for the entropy of an ideal gas.
The reference temperature of 1600~K was chosen based on the radiative-convective boundaries of Fig.~\ref{fig:jupcomp}.
For convenience, $S(3~{\rm bar},1600~{\rm K})$ was computed using the interpolated \citetalias{scvh} tables.
These include the contribution from the ideal entropy of mixing\footnote{We use the corrected version of the equations in \citetalias{scvh};
see the appendix of \citet{saumarl08}.}, a remarkably constant $S_{\rm mix} = 0.18~\Sunits$ in a large region away from H$_2$ dissociation.
Combining this $S(P,T)$ with equation (\ref{eq:Lrcb}), fixing $T=1600$~K, and
taking $\kappa(P,1600~{\rm K})\simeq0.0104(P/3~{\rm bar})^{0.2}~\mbox{cm$^2$\,g$^{-1}$}$ from the interpolated table yields
\begin{equation}
\label{eq:Lrz(M,S)}
 \log_{10} \frac{L_{\rm rz}/\LSun}{M/\MJ} = -5.05 + 1.51(S-9.6),
\end{equation}
i.e.\ $L=Mf_{\rm rz}(S)$ with $f_{\rm rz}(S)= 3.4\times10^{28}~{\rm erg~s}^{-1}\MJ^{-1} \, 10^{1.51(S-9.6)}$.
(Thus, if $\kappa\propto P^n$ and ${\rm d}S/{\rm d}\ln P=-b$,
$\log_{10} f_{\rm rz}(S) \propto (n+1)/b\ln 10$.) 
The subscript `rz' highlights that the solution applies  %
when the radiative-zero solution is reached. This fits excellently \Ae{(being mainly only 0.1~dex too high in $\log L$)}
the $L/M$--$S$ relationship found in our grid of models at intermediate entropies.

\subsubsection[High-entropy regime]{High-entropy regime: \ensuremath{L\propto M^{0.3}f_{\rm high}(S)}}
\label{sec:Lscalinghigh}
For $S\ga 10$--9 (at 3--10~$\MJ$, respectively), the luminosity becomes almost independent of mass at a given entropy.
This indicates that the radiative solution does not hold anymore, and indeed Fig.~\ref{fig:jupcomp} shows that
planets with high entropy have atmospheres extending only over a small pressure range,
with the more massive objects fully convective from the centre to the photosphere.
The shortness of the atmosphere is due to the opacity's rapid increase inward, as constant-$\kappa$ contours are almost
perpendicular to $T(P)$ profiles in that region. As in Section~\ref{sec:Lscalinglow}, we look at the behaviour of
$T$, $P$, $\kappa$, and $\delad$ at the RCB as a function of $M$ and $S$.

Fig.~\ref{fig:jupcomp} shows that at a fixed mass, $\Trcb$ is almost independent of the (high) entropy,
with the actual scaling closer to $\Trcb^4 \propto 10^{0.2S}$. Also,
$\kappa_{\rm RCB}$ is again mostly independent of $S$, as for the low entropies.
At high entropies, $\delad_{\rm RCB}$ drops continuously with increasing $S$ (decreasing $\Prcb$),   %
with a very rough $\delad \propto10^{-0.13S}$. For the fourth relation, we can fit $\Prcb\propto10^{-0.6S}$.
Combining all this with equation (\ref{eq:Lrcb}) then gives $L\propto 10^{0.7S}$, which is quite close
to a fit $L\propto10^{0.6S}$.

At fixed high entropy, $\delad$ is somewhat constant at $0.25$--0.13 for $S=10.5$--12.
Also, above $\simeq1500$~K and at $P<1$~bar, the opacity is approximately independent
of pressure, scaling only with temperature as $\kappa\sim T^4$.  %
Finally, as in the low-entropy regime, $\Prcb\propto M^{0.5}$; combining the four relations,
we should have $L\propto M^a$ with $a\simeq 0.5$--0.4.
This is not far from the direct fit $L\propto M^{0.3}$.

Therefore, the luminosity in the high-entropy regime is
\begin{equation}
\label{eq:Lhigh}
 L_{\rm high} = 10^{-3.88}~\LSun\, \left(\frac{M}{\MJ}\right)^{0.3}\,10^{0.6(S-10.5)}, 
\end{equation}
which defines $f_{\rm high}(S)$ up to a multiplicative constant. Fig.~\ref{fig:SL} indicates that
\Ae{the luminosity at the lower masses ($M\la2~\MJ$) depends more steeply on $S$ ($\simeq 10^{0.85\,S}$ at 1~$\MJ$)},
but this was ignored when obtaining equation (\ref{eq:Lhigh}).

Before summarising,
let us briefly digress about the $\Prcb\propto M^{0.5}$ scaling seen both at fixed low entropy
and at fixed high entropy. In both cases, the model grid shows that, as a reasonable approximation,
$\Prcb\propto P_{\rm phot}$.   %
At low entropies, $R$ is constant, such that $\Prcb\propto P_{\rm phot} = 2g/3\kappa \propto M/\kappa$.
\Ae{Then, the opacity's scaling of $\kappa\sim P$ (see Section~\ref{sec:Lscalinglow}; the exponent is actually
closer to $\simeq0.95$)} immediately implies roughly $\Prcb\propto M^{0.5}$.  %
At high entropies, planetary radii are significantly larger and vary, such that the radius dependence
of the photospheric pressure should not be neglected; thus $P_{\rm phot}\propto M/R^2\kappa$.
Fitting the $R(M, {\rm fixed~high~}S)$ relation in our models, we find $R\sim M^{-1.1}$ or $R\sim M^{-1.2}$, depending on the entropy.
Then, $\delad\simeq0.2$ and $\kappa\sim T^4$ yield $\kappa\sim P^{0.8}$ at fixed $S$,
and thus, isolating, $\Prcb\sim M^{0.6}$ or $\Prcb\sim M^{0.7}$.
This is a slightly stronger dependence on mass than what is found in the grid,
but the argument shows how the rough scaling can be derived.

\subsubsection{Summary of luminosity scalings}
\label{sec:lumscal}
In summary, we found from fitting the $L(M,S)$ relation across our planet models that
$L \propto M^a\,10^{\lambda S}$, with $a=(0.7,1,0.3)$ and $\lambda=(1.3,1.5,0.6)$
at low ($S\la9$--8, for 1--10~$\MJ$ respectively), intermediate ($S\la10.2$--9.6),
or high entropy, respectively, from considering the behaviour of the different factors in equation (\ref{eq:Lrcb}).
\Ae{Fitting directly the relations in the grid gives very similarly
\begin{subequations}
\label{eq:LMSfit}
\begin{align}
 L_{\rm low} &= 1.5\times10^{-7}~\LSun\, {\tilde M}^{0.72} 10^{1.3\,(S-8.2)},\\
 L_{\rm rz} &= 7.2\times10^{-6}~\LSun\, {\tilde M}^{0.98} 10^{1.58\,(S-9.6)},\\
 L_{\rm high} &= 8.7\times10^{-5}~\LSun\, {\tilde M}^{0.29} 10^{0.58\,(S-10.2)},
\end{align}
\end{subequations}
with $\tilde M \equiv M/\MJ$, to $\simeq0.01$~dex in $\log L$ except for $M\lesssim2~\MJ$ at high entropies.
}

Our approximate understanding of the different regimes is the following.
For the conditions found in the atmospheres of intermediate-entropy planets,
contours of constant opacity are almost parallel to adiabats,
\Ae{which is equivalent to saying that $\kappa$ is almost constant along lines of constant $R$ (see Fig.~\ref{fig:jupcomp}),
where $\log R \equiv \log \varrho/{T_6}^3$ and $T_6 = T/10^6$~K.}
Since $\nabla_{\rm rad}\propto \kappa$ by equation (\ref{eq:delrad}) and, for a constant $\nabla_{\rm ad}$, only $\nabla_{\rm rad}$
determines when the atmosphere becomes convective,
a slow inward increase of the opacity causes the radiative zone to extend over a large pressure range.
This in turn means that the atmosphere can reach the radiative-zero solution, which we have shown necessarily
implies $L=M f(S)$.
For the conditions found in the atmospheres of planets with high and low entropy, however,
opacity increases relatively quickly along an adiabat.  %
Since the $T(P)$ slope in the atmosphere is not too different
from that of the convective zone's adiabat, this means that opacity increases rapidly in the atmosphere,
which therefore cannot join on to the radiative-zero solution before becoming convective.
It is interesting to note that the transition from low to intermediate entropies is accompanied
by a `second-order' (i.e.\ relatively small) change in $a$ and $\lambda$,
while the physical explanation changes to `zeroth order'.

The different $L(M,S)$ scalings then reflect in part the approximate
temperature- and pressure-{\em in}dependence of the opacity at low and high $S$, respectively,
\Ae{and the fact that opacity increases relatively little along adiabats at intermediate $S$.   %
While developing even a rough analytical understanding of the various opacity scalings would be
interesting but outside the scope of this work, we briefly indicate the major contributors
in the high- and intermediate-$S$ regimes. (The following temperatures should all be understood
as somewhat approximate; cf.\ Fig.~\ref{fig:jupcomp}). Moving from 2800~K (or 3000~K at higher $P$),
above which continuum sources dominate,
down to 1300~K, the decrease in opacity is due to the settling of the rovibrational levels of H$_2$O.
Similarly, the settling of the rovibrational levels of CH$_4$ dominates from 800~K down to 480~K,
with H$_2$O also contributing. Around 2000~K, where opacity is nearly independent of pressure across
a wide pressure range, H$_2$O dominates the opacity, with its abundance remaining rather constant.
In the intermediate-entropy range, from 1300 or 1600~K to 800~K, H$_2$ dominates the composition
but it is the appearance of CH$_4$ which is crucial for the opacity.
The same qualitative behaviours can be found in the data of \citet{ferg05}
(J.\ Ferguson 2013, priv.\ comm., who also provided the information just presented). %
\citet{ferg05} do not include the powerful alkali Na, K, Cs, Rd, and Li as \citet{freed08} do,
which can raise the opacity by some 0.2~dex near  %
the RCB of intermediate-entropy planets.
}

\subsection{Luminosity as a function of helium fraction}
\Ae{
The standard grid used for analyses in this work uses a helium mass fraction $Y=0.25$
but results can be easily scaled to a different $Y$.
Following a suggestion by D.\ Saumon (2013, priv.\ comm.), one can write
\begin{equation}
 \left(\frac{{\rm d} L}{{\rm d}Y}\right)_{M,S} \approx \left(\frac{\partial L}{\partial \ln S}\right)_{M,Y} \times
       \left.\frac{{\rm d}\ln S}{{\rm d}Y}\right|_{\left(P_0,T_0\right)},
\end{equation}
where $(P_0,T_0)$ is some appropriate location,
and compute $\gamma S\equiv{\rm d} S/{\rm d}Y = (S_{\rm H} - S_{\rm He}) + {\rm d}S_{\rm mix}/{\rm d}Y$,
where $S = (1-Y)S_{\rm H} + YS_{\rm He} + S_{\rm mix}$, $S_{\rm H}$, $S_{\rm He}$, $S_{\rm mix}$ are respectively
the total, hydrogen, helium, and mixing entropies per baryon \citepalias{scvh}.
Given equation (\ref{eq:Lrcb}), one might heuristically expect $(P_0,T_0)$ to be
near the radiative--convective boundary (RCB), at least when determining ${\rm d}S/{\rm d}Y$ at constant $L$.
Indeed, we find this to be the case, with $\gamma\simeq -0.63$ at the RCB of the $Y=0.25$
or 0.30 models for intermediate to high entropies. Thus, for $Y'$ sufficiently close to 0.25,
\begin{equation}
X(M,S',Y') = X(M,S,Y)
\end{equation}
if
\begin{equation}
\label{eq:S'S}
S' = S \times\left[1 - 0.63\left(Y'-0.25\right)\right],
\end{equation}
where $X$ is a planet property such as $L$ or $\tau_S$. 
For $Y=0.27$, this is at worse accurate only to 0.05~dex in luminosity towards high entropies for masses below $\simeq12~\MJ$
when $S\la8$. This rescaling is also adequate (in the same domain) for $X=R$, to 15 per cent towards {\em high} entropies,
but a more accurate fit can be obtained with $\gamma\simeq-0.3\pm0.1$. Interestingly,
nowhere within a planet structure in the grid does $|\gamma|$ drop below $0.55$;
the explanation in this case (in analogy to $\gamma$ being evaluated at the RCB for $L$) is not clear.
}

\Ae{
Since at the RCBs in the grid the hydrogen is molecular, one might try to obtain $\gamma$ analytically with the Sackur--Tetrode formula.
Neglecting the subdominant contribution from the rotational degrees of freedom of H$_2$ yields
$S_{\rm H}/S_{\rm He} \approx \mu_{\rm He}/\mu_{{\rm H}_2} = 2$, where $\mu_X$ is the molecular mass of $X$.
This leads to $\gamma=-0.57$, which is not far from the accurate result $-0.63$
yet shows that the ${\rm d}S_{\rm mix}/{\rm d}Y$ term cannot be neglected.
}

\Ae{Equation (\ref{eq:S'S}) makes it simple to convert e.g.\ constraints on the initial entropy based on
cooling curves with a particular $Y$ to another set with a different $Y$, regardless of the approach used for the cooling.
The typically small change in the entropy ($\simeq3$ per cent) is nevertheless significant because of the strong dependence of $L$ on $S$.
}

\subsection{Cooling}
\label{sec:gencooling}
\Ae{One can also derive the functional form of the cooling tracks with a few simple arguments.}
With $L \propto M^a\,10^{\lambda S}$,
equation (\ref{eq:L}) implies ${\rm d}S/{\rm d}t = -M^{a-1}f(S)/\overline{T}$.
Since hydrostatic balance yields $\Pc \sim GM^2/R^4$ and the convective core is adiabatic,
we expect $\overline{T}\simeq \Tc\propto (M^2/R^4)^{\delad}$. (Across the whole grid, $\overline{T}/\Tc = 0.55$--0.63.)
Therefore, ignoring the radius dependence \Ae{(since $4\delad$ is not large and $R$ is rather constant at lower entropies)},
\Ae{$-{\rm d}S/f(S) \propto M^{-2\delad-1+a}\,{\rm d}t$}.
This means that the entropy of a cooling planet should be approximately a function of $t/M^{2\delad\Ae{+1-a}}$,
and that more massive planets cool more slowly since $a$ is always smaller than $2\delad+1\simeq5/3$.
\Ae{In fact, using that $f(S)=10^{\lambda S}$, one can compute the integral to obtain an analytic expression for the cooling tracks:
\begin{equation}
\label{eq:cool}
 \frac{1}{L(t)} = \frac{1}{\Li} + \frac{1}{L_{\rm hs}(t)}, \;\;\; L_{\rm hs}(t) =  \frac{M^{2\delad+1}}{\beta\, t},
\end{equation}
where $\Li$ and $L_{\rm hs}$ are the initial and hot-start luminosities, respectively,
and $\beta=C\lambda\ln10$ with $C$ a dimensional constant grouping prefactors. We have assumed that
$a$, $\lambda$, and $\delad$ do not change as the planet cools.}
Thus, $L\propto M^a$ at fixed entropy, with $a=0.3$--1,
but at a fixed time the luminosity has a steeper dependence on $M$,
\Ae{$L\propto M^{2\delad+1} \simeq L\propto M^2$ for intermediate entropies}.
A more detailed analytic understanding of the cooling curves
for irradiated planets along these lines was developed by \citet{ab06},
\Ae{and a careful, approximate but surprisingly accurate analysis for brown dwarfs
may be found in \citet{burr93}.}

\Ae{
Equation (\ref{eq:cool}) approximately describes cooling tracks with arbitrary initial entropy,
with our cooling curves well fitted (to roughly 0.1~dex) below $10^{-5}~\LSun$ by
$C=3\times10^{-2}$~cgs when $\lambda=1.5$ and $\delad=1/3$.
At higher luminosities, in particular for hot starts, a better fit (with the restriction of $L\ga10^{-6}~\LSun$ if $M\la2.5~\MJ$)
which captures the average shape and spacing of the cooling curves is provided directly by the classical result of \citet{burr93}, who find
\begin{equation}
\label{eq:hs_BL93}
 L_{\rm hs} = 7.85\times10^{-6}~\LSun\,\frac{\left(M/3~\MJ\right)^{2.641}}{\left(t/10~\textrm{Myr}\right)^{1.297}}
\end{equation}
(for $\kappa=0.01$~cm$^2\,$g$^{-1}$), i.e.\ a somewhat different mass and time dependence.
}

\subsection{Comparison with classical hot starts \Ae{and other work}}
\label{sec:modelcomp}
\begin{figure}
\includegraphics[width=84mm]{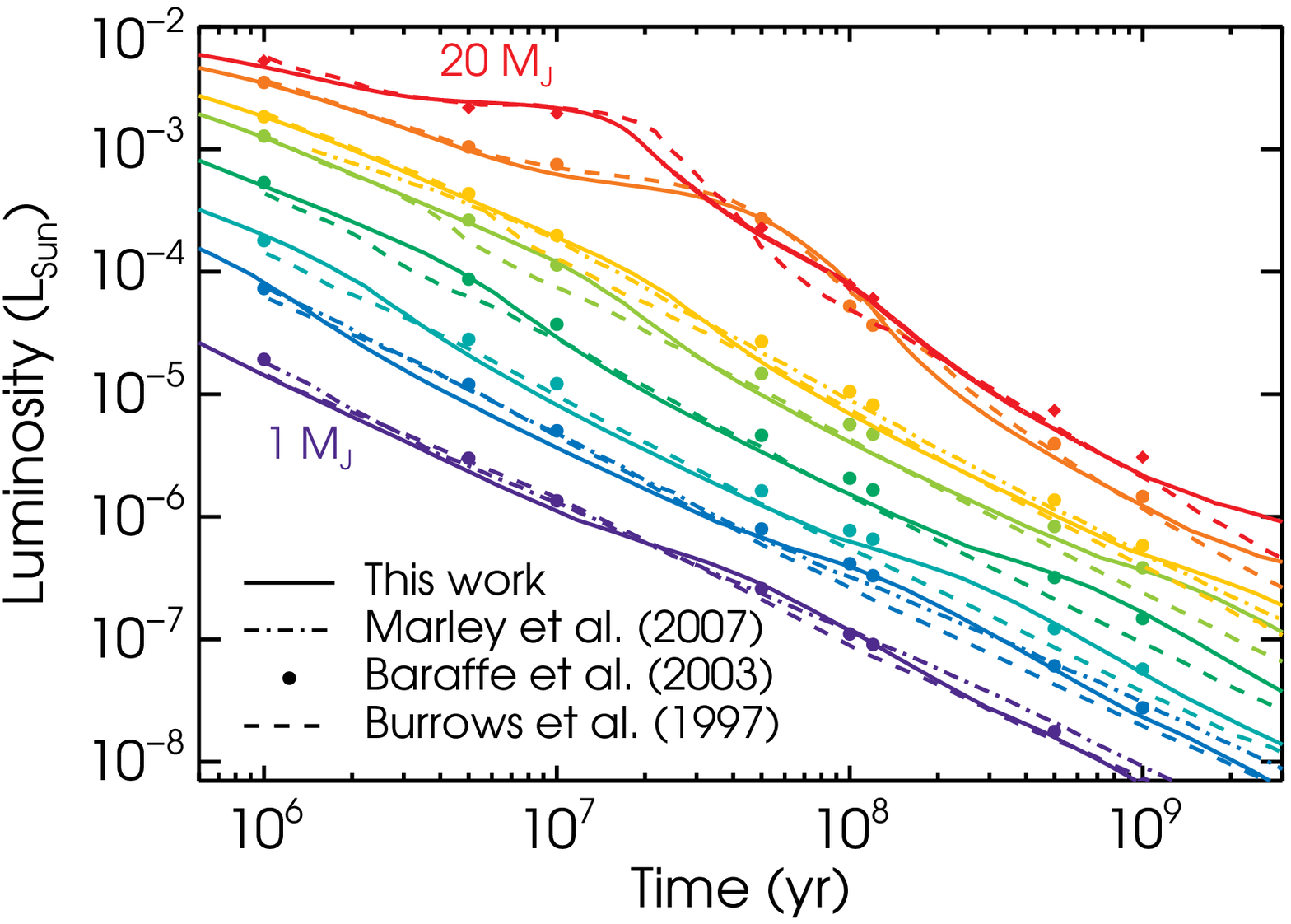}
\caption{Cooling curves for (bottom to top) $M=1,2,3,5,8,10,15,$ and $20~\MJ$ (solid curves) compared with
the \citet{marl07} hot starts (for $M=1,2,$ and 10~$\MJ$ only; dash-dotted curves), the COND03 tracks (dots)
and the \citet{burr97} models (dashed curves). The COND03 data is shown as dots because
of insufficient sampling of the cooling tracks at higher masses.
}
\label{fig:hotcoolingcurves}
\end{figure}

We now compare our cooling curves to classical hot starts.
Fig.~\ref{fig:hotcoolingcurves} shows cooling curves for large initial entropy compared to the hot-start models
of \citet{marl07}, the COND03 models \citep{bara03}
and those of \citet{burr97}, all of which use non-grey atmospheres with detailed opacities.
The agreement is excellent, with our luminosities within the first 3~Gyr approximately within $-30$ and 20 per cent and $-30$ and 60 per cent 
above those of \citeauthor{bara03} and \citeauthor{burr97}.
(Of interest for the example of Section~\ref{sec:gTconstr}, our radii along the cooling sequence are
at most approximately \Ae{two to five} per cent greater \Ae{at a given time}. %
This difference is comparable to the effect of neglecting heavy elements in the equation of state (EOS) or
not including a solid core \citep{saumon96} and not significant for our purposes.
\Ae{See also Appendix~\ref{app:R}.}) %
Our deuterium-burning phase at 20~$\MJ$ ends slightly earlier than in \citet{burr97} but this might be due to our
simplified treatment of the screening factor.

\begin{figure}
\includegraphics[width=84mm]{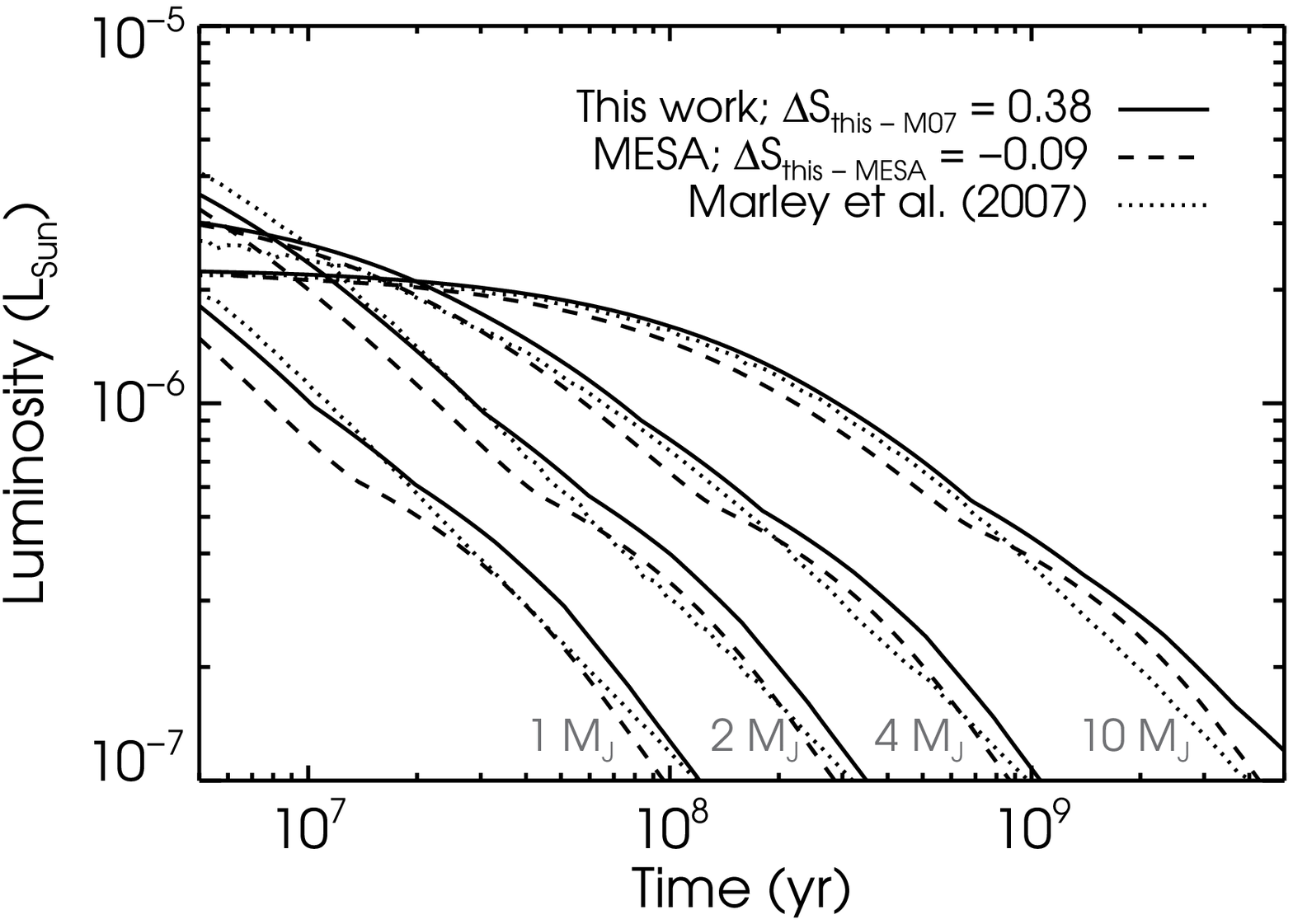}\\
\includegraphics[width=84mm]{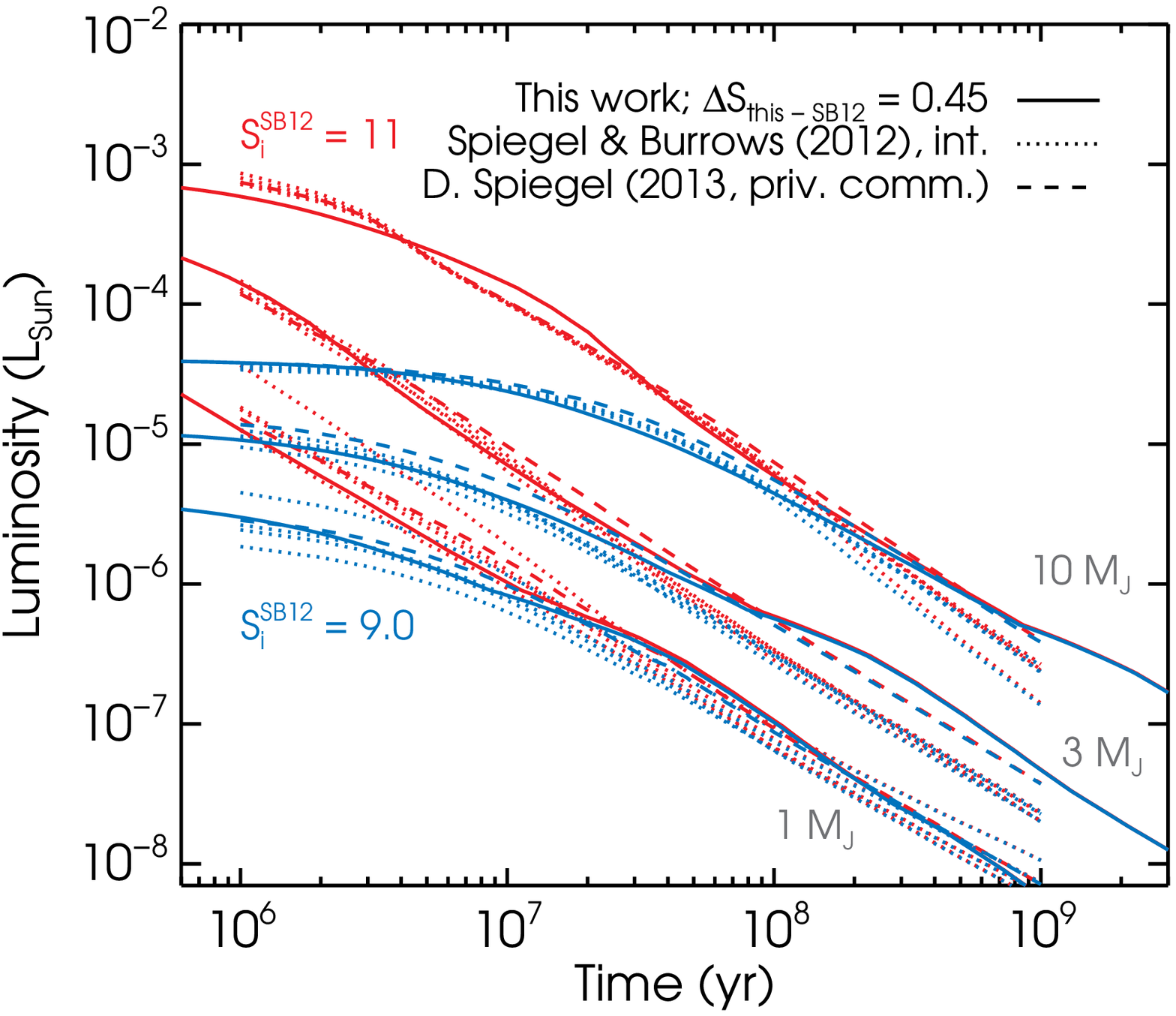}
\caption{
{\em Top panel:}
Cooling curves for $M=1,2,4,10$~$\MJ$ and initial entropies equal to the \citet[M07]{marl07} `tuning-fork diagram' (their fig.~2)
values $\Si=9.23,9.00,8.60,8.23$. \Ae{The initial entropies for our cooling tracks were increased by 0.38~$\Sunits$
to compare with \citet{marl07}, with those of \textsc{mesa} further increased by 0.09~$\Sunits$.}
{\em Bottom panel:}
Cooling curves for $M=1,3,10~\MJ$ (bottom to top) with initial entropies $\Si=9.0$
and $\Si=11$ increased by 0.45~$\Sunits$ \Ae{mainly for `thermodynamic reasons'} (see text; solid lines),
compared to those of \citet[SB12]{spiegel12} with $\Si=9.0$ and $\Si=11$ (dashed and dotted lines).
\Ae{For each mass and initial entropy, the four dotted lines (some overlap) correspond to the integrated spectra
for the four atmosphere types in no particular order, serving only to indicate a range.}   %
The red-side truncation of the \citet{spiegel12} spectra causes an underestimation of the bolometric luminosity at late times (see text).
\Ae{The dashed line at each mass and initial entropy shows the bolometric luminosity directly from the models.}
}
\label{fig:coolingcurvesSpiegel}
\end{figure}

Fig.~\ref{fig:coolingcurvesSpiegel} shows cooling curves for lower initial entropies than in Fig.~\ref{fig:hotcoolingcurves}.
The cooling curves show the behaviour found by \citetalias{marl07} and \citet{spiegel12} in which the luminosity initially varies very slowly, with the cooling time at the initial entropy much larger than the age of the planet. Eventually, the cooling curve joins the hot start cooling curve once the cooling time becomes comparable to the age.

Comparing to the cold starts in fig.~4 of \citetalias{marl07}, our models are a factor of $\simeq3.5$--3.9 lower in luminosity for the same
initial entropies, as given in their fig.~2.
Increasing our initial entropies by 0.38~$\Sunits$ brings our cooling curves into agreement with theirs when the planet has not yet
started cooling. (For this comparison, we do not correct the time offset for the higher masses (see figs.~2 and 4 of \citetalias{marl07}),
which are already on hot-start cooling curves at the earliest times shown.) \Ae{As mentioned in Appendix~\ref{sec:Sdiff},
this implies a real difference between our $L(S)$ of only 0.14~$\Sunits$.}

\citet{spiegel12} computed the evolution of gas giants starting with a wide range of initial entropies.
To compute the bolometric luminosity of their models, we take the published spectra and integrate the flux in the wavelength
range given, 0.8--15~$\umu$m. The bottom panel of Fig.~\ref{fig:coolingcurvesSpiegel} shows the comparison
to all four model types, with or without clouds and at solar or three times solar metallicity.
Increasing our entropy by 0.45~$\Sunits$ -- e.g.\ comparing the \citet{spiegel12} model
with $S=9.0$ to ours with $S=9.45$ -- yields very good agreement, with our luminosities overlying their curves
or within the spread due to the different atmospheres. \Ae{As discussed in Appendix~\ref{sec:Sdiff},
this is mainly due to a constant entropy offset of 0.52~$\Sunits$ between the tables used by the \citeauthor{burr97} group
and the published \citetalias{scvh} tables used in the present work, leaving a net offset of merely 0.07~$\Sunits$.}

The apparent disagreement with \citet{spiegel12} at late times comes from the increasing fraction of the flux in the Rayleigh--Jeans tail
of the spectrum beyond 15~$\umu$m. For comparison, the implied required bolometric correction is equal to 10 to 50 per cent of the flux
in 0.8--15~$\umu$m for a blackbody with $\Teff\simeq700$--300~K. From this and the hot-start $\Teff$ tracks shown in \citet{spiegel12},
we estimate that integrating the spectrum should give a reasonable estimate (to ca.\ 30 per cent)
of the bolometric luminosity only up to $\simeq50$, 200, and 1000~Myr for objects with $M=1$, 3, and 10~$\MJ$ respectively.
This is indeed seen in Fig.~\ref{fig:coolingcurvesSpiegel}.
\Ae{Bolometric luminosities kindly provided by the authors (D.\ Spiegel 2013, priv.\ comm.), which are rather insensitive
to the atmosphere type, are also shown for a more direct comparison
and confirm the reasonable match of our cooling curves with those of \citet{spiegel12}.}

\Ae{We have also computed cooling curves} with the \textsc{mesa} stellar evolution code \citep[][\Ae{revision 4723}]{paxton11,paxton13},
\Ae{and they are in excellent agreement with our results.}
We compared our $L(S)$ relation to ones obtained from \textsc{mesa}, \Ae{(also with $Y=0.25$)} at different masses  %
and found that they are very nearly the same, with an entropy offset $\Delta S\la0.1$.  %
\Ae{
Moreover, Fig.~\ref{fig:coolingcurvesSpiegel} shows that the agreement of the time evolution is quite good,
with in particular the late-time `bumps' due to opacity when the cooling curves enter the intermediate-entropy regime
(cf.\ Figs.~\ref{fig:jupcomp}, \ref{fig:SL}, and~\ref{fig:coolingS}).
}
We also produced grids with other opacities, using the \textsc{mesa} tables with the default $Y=0.28$ and $Z=0.02$ \Ae{(for the opacity calculation only)},
and the \citet{freed08} tables with $[{\rm M/H}]=\pm0.3$~dex, and found that this   %
changed the luminosity by at most $\simeq10$ per cent
\Ae{
at a given mass and internal entropy.
Similarly, small differences were found to result from a changed helium mass fraction in the bulk of the planet at a given mass
and entropy per \emph{nucleus}.  %
}

The upshot of these comparisons to classical, non-grey-atmosphere hot and cold starts is that
cooling tracks computed with the simple and numerically swift cooling approach described above can reproduce models
which explicitly calculate the time dependence of the luminosity.
When comparing models from different groups, one should keep in mind that there \Ae{can be a systematic offset} in the entropy
values of $\DSsyst =0.52$ \Ae{(at $Y=0.25$) due to different versions of the \citetalias{scvh} EOS, which however has no physical consequence for the cooling}.
\Ae{Moreover, the remaining difference $|\Delta S| \la 0.15$ is} small compared to the entropy \Ae{range} between hot and cold starts.

\section{General constraints from luminosity measurements}
\label{sec:genconstr}

Masses of directly-detected exoplanets are usually inferred by fitting hot-start cooling curves
\citep[e.g.][]{burr97,bara03} to the measured luminosity
of the planet, using the stellar age as the cooling time. Since the hot-start luminosity
at a given age is a function only of the planet mass, the measured luminosity {determines} the planet mass.
\Ae{Equation (\ref{eq:hs_BL93}) provides a quick estimate of this `hot-start mass':
\begin{equation}
\label{eq:Mhs}
 M_{\rm hs} = 3~\MJ\,\left(\frac{L}{7.85\times10^{-6}~\LSun}\right)^{0.379}\left(\frac{t}{10~\mbox{Myr}}\right)^{0.491}.
\end{equation}
Moreover}, a planet's luminosity, at a given time and for a given mass, can never exceed that of the hot starts,
since a larger initial entropy would have merely cooled on to the hot-start cooling track at an earlier age.

However, we have seen above that the luminosity at a given mass can be lowered by considering a sufficiently smaller
initial entropy, which might be the outcome of more realistic formation scenarios \citep{marl07,spiegel12}.
With the fact that luminosity increases with planet mass \Ae{at a given entropy}, this simple statement has important consequences for the
interpretation of direct-detection measurements, namely that %
{\em there is not a unique mass which has a given luminosity at a given age}.
Cold-start solutions correspond to planets not having forgotten their initial conditions, specifically their initial
entropy $\Si$, and every different initial entropy is associated to a different mass.
In other words, a point in $(t,L)$ space -- a single brightness measurement -- is mapped to a curve in $(M,\Si)$ space.
Since \citet{marl07} \citep[but see also][]{bara02,fort05},  %
it is generally recognised that direct detections should not be interpreted to yield a unique mass solution,    %
but, with the exception of \citet{bonn13}, who used infrared photometry, this is the first time
that this degeneracy is calculated explicitly.

\subsection[Shape of the M-Si constraints]{Shape of the \ensuremath{\bmath{M}}--\ensuremath{\bmath{\Si}} constraints}
\label{sec:MSishape}
\begin{figure}
\includegraphics[width=84mm]{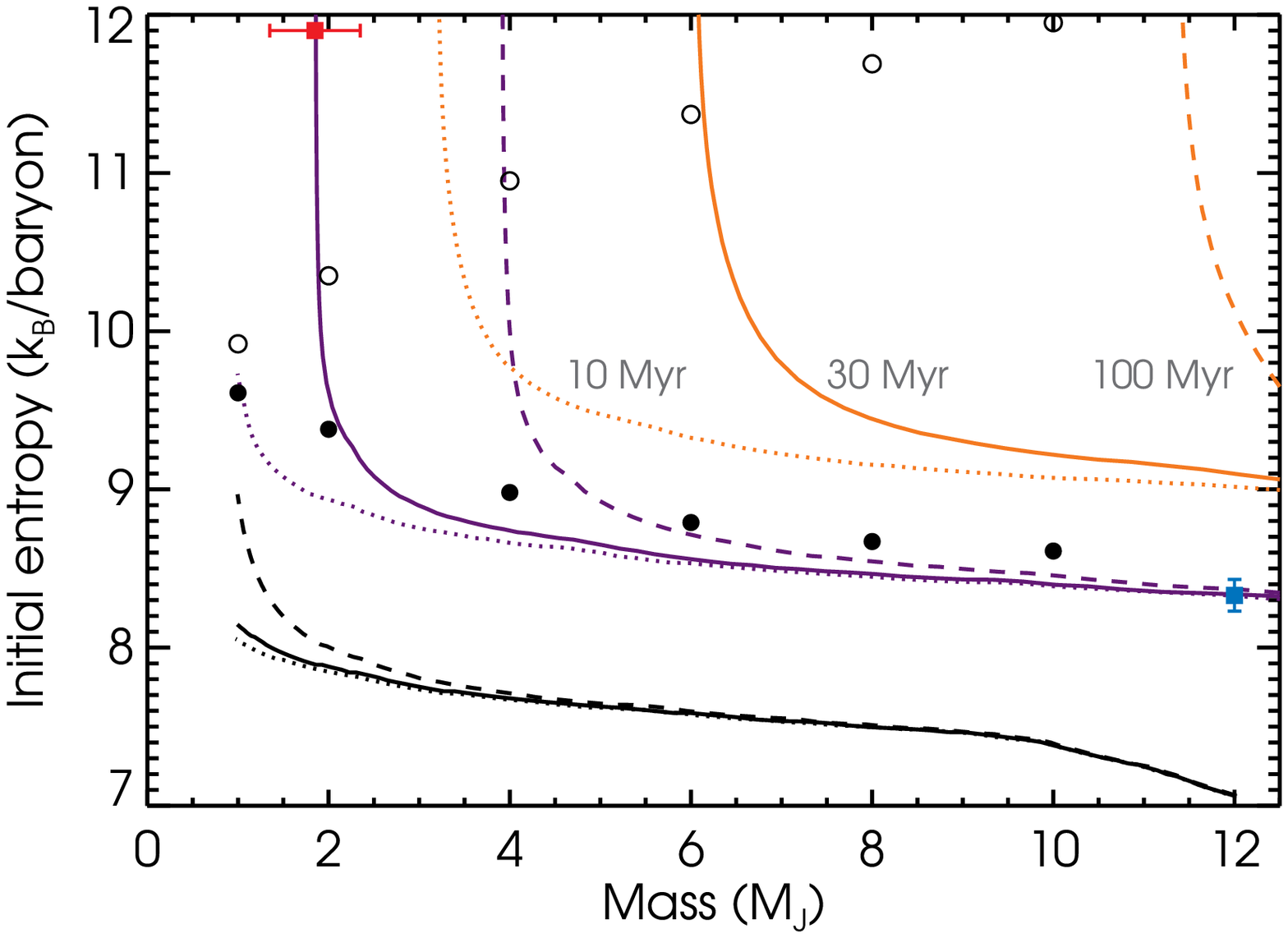}\\
\includegraphics[width=84mm]{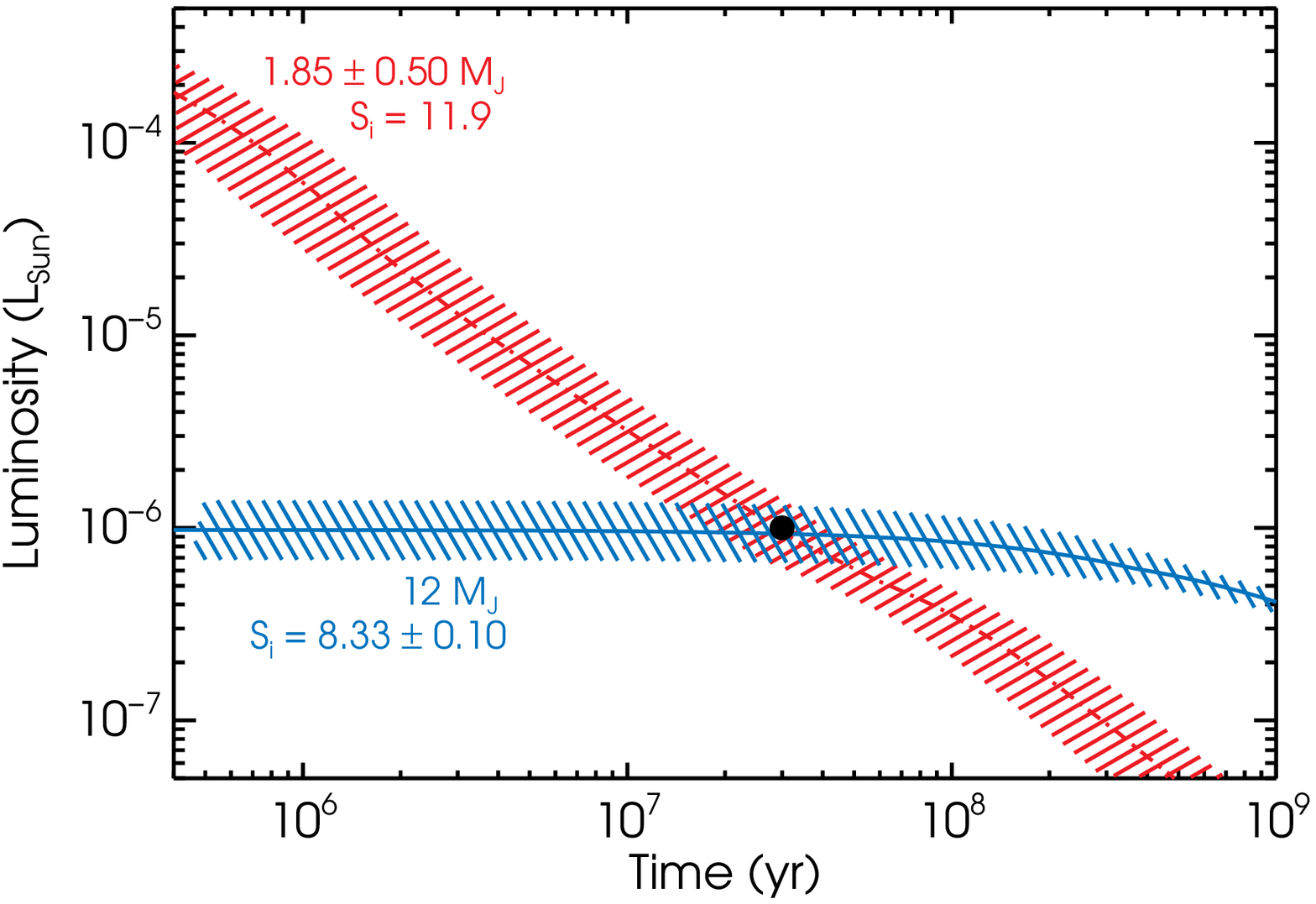}
\caption{
{\em Top panel:}
Allowed values of planet mass $M$ and initial entropy $\Si$ corresponding to $\Lbol=10^{-7}$, $10^{-6}$, and $10^{-5}~\LSun$
(bottom to top sets of curves) and ages $t=10$, $30$, and 100~Myr (left to right or dotted, full and dashed lines).
The circles show the results from \citet{marl07}, increased by 0.38~$\Sunits$ (cf.\ Fig.~\ref{fig:coolingcurvesSpiegel}),
for cold starts (filled circles) and hot starts (open circles), where the entropy is 1~Myr after the onset of cooling.
Red and blue squares indicate
the illustrative $M$ and $\Si$ used for the lightcurves in the bottom panel, with errorbars corresponding to the hatched regions there.
{\em Bottom panel:}
Two examples of cooling curves that have $L=10^{-6}~\LSun$ at 30~Myr. The dot-dashed (red) line has $M=1.85~\MJ$ and $\Si=11.9$
(hot start), while the full (blue) one has $M=12~\MJ$ and $\Si=8.33$ (cold start). Hashed regions indicate a spread
of $0.5~\MJ$ ($0.1~\Sunits$) for the hot- (cold-)start curve.
The respective relative uncertainties are $\Delta M/M\simeq \frac{1}{2}\Delta t/t$ and $\Delta\Si \simeq0.7\Delta\log_{10}\Lbol$
for the cold- and hot-start cases, respectively (see text).
}
\label{fig:MSplotnoD}
\end{figure}

The top panel of Fig.~\ref{fig:MSplotnoD} shows the allowed masses and initial entropies for different values of luminosity
$\Lbol/\LSun=10^{-7}$, $10^{-6}$ and $10^{-5}$, at ages of $10$, $30$, and $100$~Myr.
Below the deuterium-burning mass, constant-luminosity curves in the $M$--$\Si$ plane have two
regimes. At high initial entropies, the derived mass is the hot-start mass independent of $\Si$ since all $\Si$ greater
than a certain value have cooling times shorter than the age of the system.
There, uncertainty in the stellar age translates directly into uncertainty in the planet mass:
since $L\propto 1/t$ and $L\sim M^{2}$, the mass uncertainty is $\Delta M/M\simeq \frac{1}{2}\Delta t/t$.  %
At lower entropies, the luminosity measurement occurs during the early, almost constant-luminosity %
evolution phase. %
Given a luminosity match in this region, one can obtain another by assuming a lower (higher) initial entropy
and compensating by increasing (decreasing) the mass. As seen in Section~\ref{sec:struct+cool},
$L$ is a very sensitive function of $S$ at low and intermediate entropies, so that a small decrease in initial entropy must be compensated
by a large increase in mass to yield the same luminosity at a given time; this yields the approximately flat portion of the curves.
As long as the cooling time for a range of masses and entropies remains shorter than the age,
{\em the entropy constraints do not significantly depend on the age}.
The uncertainty in the initial entropy is $\Delta \Si \simeq 1/\lambda\,\Delta\log_{10}\Lbol$,
where $1/\lambda\simeq0.7$ or 1.7 at low or high entropy (see Section~\ref{sec:lumscal}). %

The circles in the top panel of Fig.~\ref{fig:MSplotnoD} show the initial entropies for
cold- and hot-start models from \citetalias{marl07}, the `tuning-fork diagram of exoplanets'.
The entropies were increased by 0.38~$\Sunits$ as in the top panel of Fig.~\ref{fig:coolingcurvesSpiegel}
to match our models.
Since the cooling time increases with mass (see Section~\ref{sec:gencooling}), %
heavier planets of the hot-start (upper) branch, i.e.\ with arbitrarily high initial luminosity,  %
have cooled less and are therefore at higher $S$. %
For their part, the cold-start entropies, which are still the post-formation ones,
lie close to a curve of constant luminosity $L\simeq 2\times10^{-6}~\LSun$. %
This reflects the fact that the post-formation luminosities in \citetalias{marl07}, as seen in their fig.~3,
have similar values for all masses. The cause of this `coincidence' is presently not clear; it might be a physical process
or an artefact of the procedure used to form planets of different masses.
Putting uncertainties in the precise values aside, the two prongs of the tuning fork in Fig.~\ref{fig:MSplotnoD} give
an approximate bracket within which just-formed planets might be found.

Mass information for a directly-detected planet can put useful constraints on its initial entropy
and also potentially on its age and luminosity simultaneously.
For instance, dynamical-stability analyses and radial-velocity observations (see Section~\ref{sec:obsobj}
\Ae{and Appendix~\ref{app}})
typically provide upper bounds on the masses. Since $\Si$ decreases monotonically with $M$
at a fixed luminosity, this translates into a {\em lower limit} on the initial entropy.
This has the potential of excluding the coldest-start formation scenarios.
Conversely, a lower limit on the planet's mass implies an {\em upper} bound on $\Si$.
If it is greater than the hot-start mass, this lower limit on the mass of a planet would be very powerful,
due to the verticalness of the hot-star branch.
Combined with the flatness of the `cold branch'
of the $M(\Si)$ curve, this could easily restrict the initial entropy to a dramatically small
$\Delta \Si\simeq 0.5 \sim \DSsyst$.
Also, the top panel of Fig.~\ref{fig:MSplotnoD} shows that not all age and luminosity combinations
are consistent with a given mass upper limit.
Given the often important uncertainties in the age and the bolometric luminosity, this may represent a valuable input.

\subsection{Solutions on the hot- vs.\ cold-start branch}
The bottom panel of Fig.~\ref{fig:MSplotnoD} shows lightcurves illustrating the two regimes of the $M(\Si)$
curves discussed above.
The hot-start mass is $1.85~\MJ$, whereas the selected cold-start case $(\Si=8.33)$ has $M=12~\MJ$ -- 
i.e.\ {\em a six times larger mass} --
and both reach $\log \Lbol/\LSun=-6$ at $t=30$~Myr, with the cold-start values essentially
independent of age.
The hatched region around the 1.85-$\MJ$ curve comes from hot-start solutions between 1.35 and 2.35$~\MJ$ and is within
a factor of two of the target luminosity, showing the moderate sensitivity of the cooling curves to the mass.
However, in the cold-start phase, a variation by a factor of two can also be obtained by varying at a fixed mass
the initial entropy from 8.23 to 8.43~$\Sunits$. This great sensitivity implies that 
combining a luminosity measurement with information on the mass would yield,
if some of the hot-start masses can be excluded, tight constraints on the initial entropy.

\begin{figure}
\includegraphics[width=84mm]{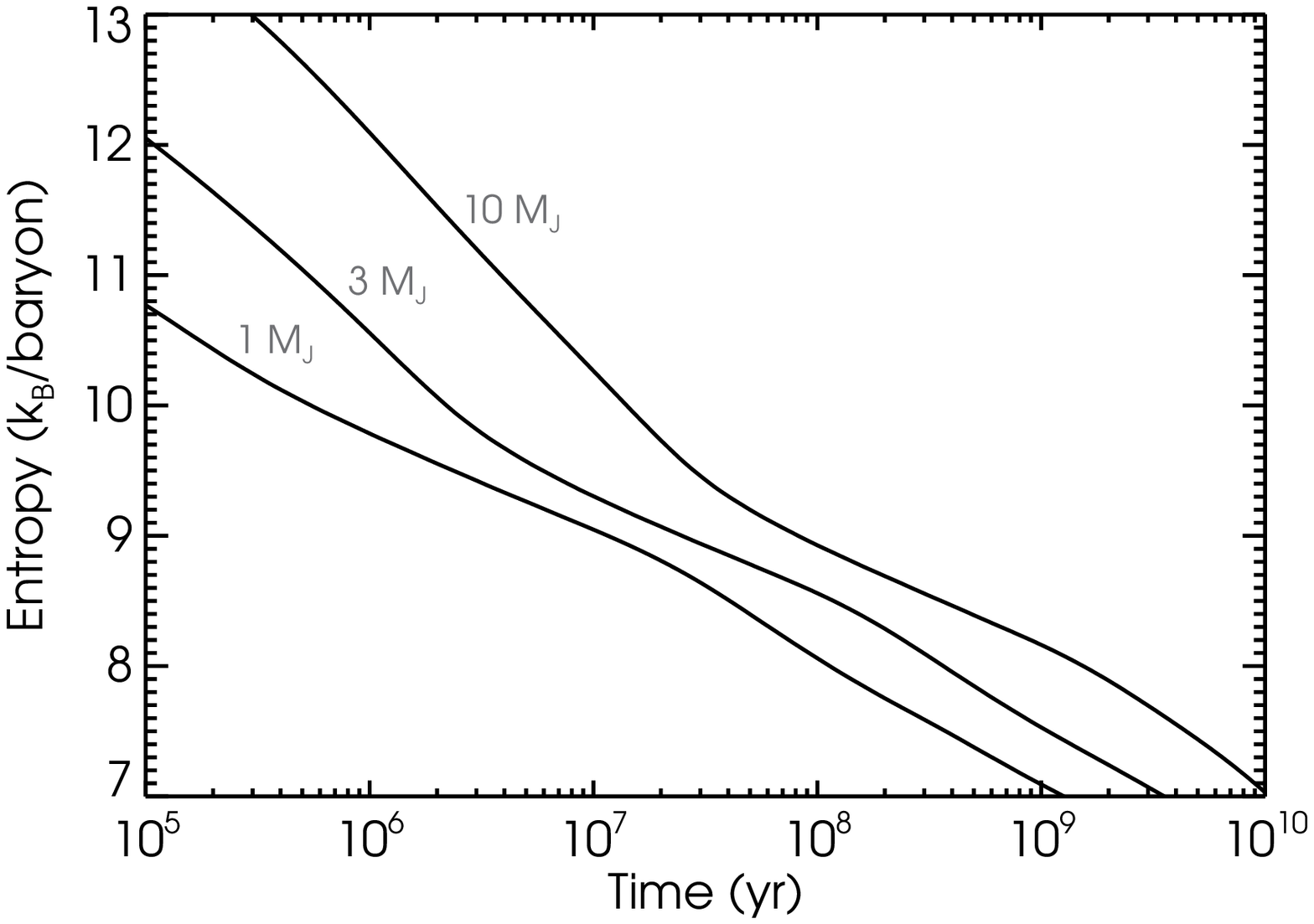}
\caption{
Entropy of hot starts as a function of time for planet masses of 1, 3, and 10~$\MJ$ (bottom to top).
At a given age, the curve indicates the value of initial entropy above which the `hot-start mass' applies.
For a planet mass larger than the hot-start value, the initial entropy must be lower than the hot-start entropy at
the current age.
}
\label{fig:coolingS}
\end{figure}

\subsection{Definition of `hot-start mass'}

By showing the entropy of hot starts as a function of time,
Fig.~\ref{fig:coolingS} provides another way of looking at Fig.~\ref{fig:MSplotnoD}. %
Given a mass obtained from hot-start cooling tracks, the entropy value at that time read off from the curves
indicates what `hot' is, i.e.\ provides a lower bound on the initial entropy if the hot-start mass is the true
mass. If however the planet is more massive, this entropy value is instead an upper bound on the post-formation entropy.
\Ae{As a rule of thumb, the entropy slope is $\simeq-2$ or $-1~\Sunits$ per time decade at early or late times, approximately,  %
with the break coming from the change in the entropy regime (cf.\ Fig.~\ref{fig:SL}).}

\section{General constraints from gravity and effective-temperature measurements}
\label{sec:gTconstr}
Before applying the analysis described in the previous section to observed systems, it is worth discussing
a second way by which constraints on the mass and initial entropy can be obtained.
The idea is to firstly derive an object's effective temperature and surface gravity
by fitting its photometry and spectra with atmosphere models. Integrating the best-fitting model spectrum gives
the luminosity, and this can be combined with $\Teff$ and $\log g$ to yield the radius and the mass.
This procedure was described and carefully applied by \citet{mohanty04a} and \citet*{mohanty04b}.
However, one can go further: considering models coupling detailed atmospheres with interiors
at an arbitrary entropy, the mass and radius translate into a mass and current entropy.
Then, using cooling tracks beginning with a range of initial conditions,
the initial state of the object can be deduced given the age.
Thus, in contrast to the case when only luminosity is used,
$M$ and $\Si$ can both be determined without any degeneracy between the two.  %

In practice, however, there seems to be too much uncertainty in atmosphere models for this method to be currently reliable,
as the work of \citeauthor{mohanty04a} shows. Their sample comprised a dozen young ($\sim 5$~Myr) objects
with $\Teff \simeq 2600$--2900~K\footnote{The spectral types are $\simeq\textrm{M}$--$\textrm{M}7.5$,
but \citet*{mohanty04b} stress that the correspondence between the spectral type    %
and effective temperature of young objects has not yet been empirically established
and thus that calibration work (in continuation of theirs) remains to be done.} with high-resolution optical spectra.
The combined presence of a gravity-sensitive Na~{\sc i} doublet
and effective-temperature-sensitive TiO band near 0.8~$\umu$m allowed a relatively precise determination
of $\log g$ and $\Teff$ for most objects, with uncertainties of 0.25~dex and 150~K \citep{mohanty04a}.
However, there were significant offsets in the gravity ($\simeq0.5$~dex) of the two coldest objects %
with respect to model predictions of \citet{bara98} and \citet{chab00}.
\citet{mohanty04a} came to the conclusion that the models'   %
treatment of deuterium burning, convection\footnote{Qualitatively, their finding that theoretical tracks predict too quick cooling at low masses
might be evidence for the argument of \citet{leconte12} that convection in the interior of these objects could
be less efficient than usually assumed.}  %
or accretion -- i.e.\ the assumed initial conditions -- are most likely responsible for this disagreement at lower $\Teff$.
Moreover, the more recent work of \citet{barman11_1207} (see Section~\ref{sec:1207_tL} below) indicates that `unexpected' cloud thickness
and non-equilibrium chemistry may compromise a straightforward intepretation of spectra in terms of gravity and temperature
for young, low-mass objects. \Ae{(See also \citet{moses13} for a review of photochemistry and transport-induced quenching
in cool exoplanet atmospheres.)} Nevertheless, with the hope that future observations will allow
a reliable calibration of atmosphere models, we illustrate with an example how $M$ and $\Si$ can be determined
for an object from its measured $\Teff$ and $\log g$.

\begin{figure}
\includegraphics[width=84mm]{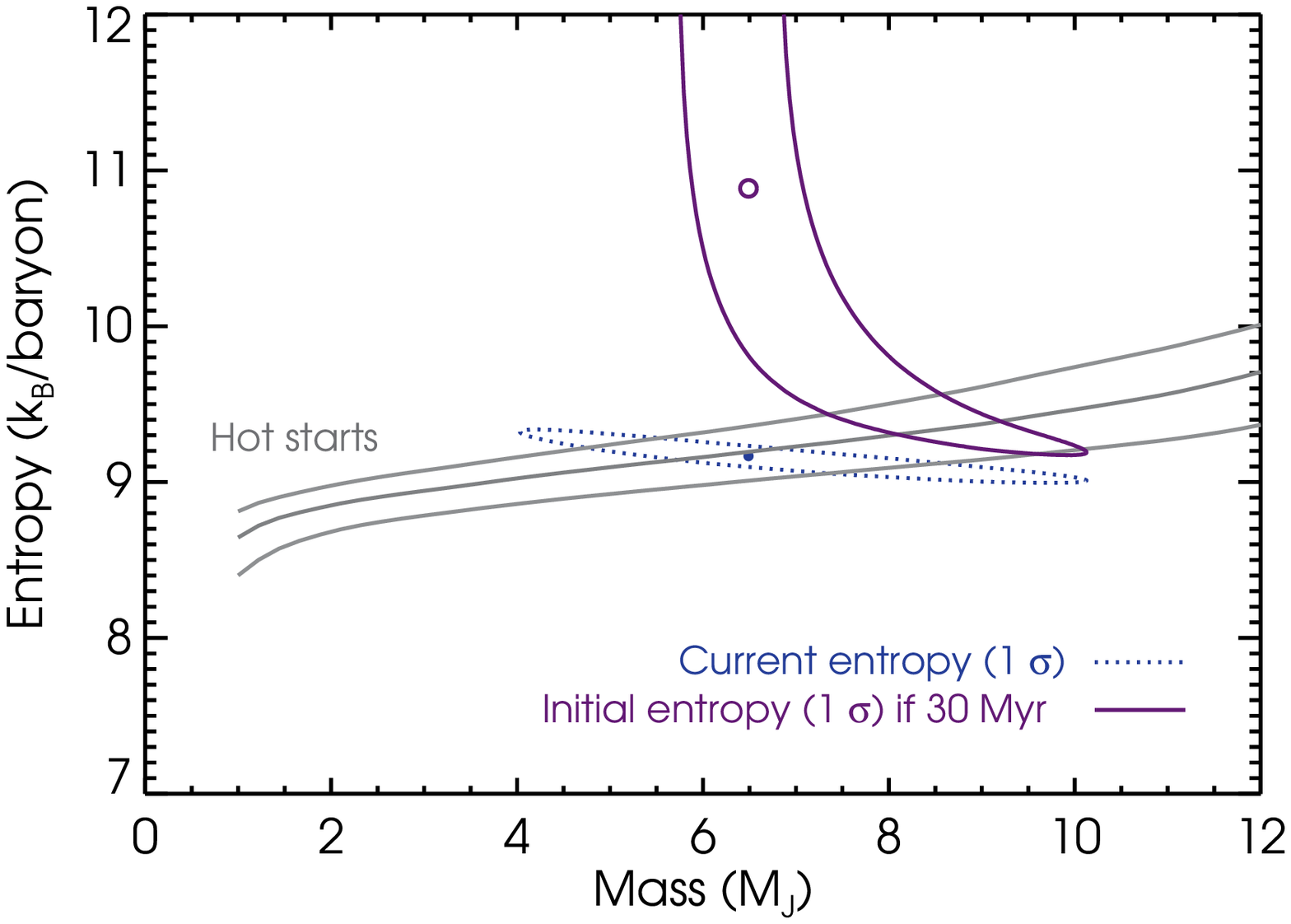}
\caption{
Constraints on the mass and entropy of an example planet from its gravity and effective temperature,
with $\log g=4.00\pm0.25$~\mbox{(cm\,s$^{-2}$)} and $\Teff=900\pm50$~K (and thus optimistic errorbars; see text).
The mass and current entropy corresponding to $\log g=4.00$ and $\Teff=900$~K are shown by a filled circle.
The narrow ellipse (\Ae{dotted} line) shows the 1-$\sigma$ confidence region in the mass and {current} entropy,
whereas the deformed ellipse (\Ae{solid} line) shows the constraints on the {initial} entropy assuming an age of 30~Myr.
The open circle corresponds to the `measured' central value.
The three solid grey lines indicate the entropy of hot starts at 20, 30, and 50~Myr (top to bottom).
Higher masses can be seen to cool more slowly (see Section~\ref{sec:gencooling}).
}
\label{fig:MSloggT_eg}
\end{figure}

Fig.~\ref{fig:MSloggT_eg} shows the constraints on the current mass and entropy of a planet with
$\log g=4.00\pm0.25$~\mbox{(cm\,s$^{-2}$)}
and $\Teff=900\pm50$~K,
where the uncertainties in the gravity and effective temperature are the possible accuracy reported by \citet{mohanty04a}
and thus correspond to an optimistic scenario.
The constraints are obtained by simultaneously solving for $M$, $S$, $R$, and $L$ given $\Teff$ and $\log g$,
with the $L(M,S)$ and $R(M,S)$ relations given by our model grid. %
The required mass and current entropy are $M=6.5~\MJ$ and $S=9.2$, with the 1-$\sigma$ ellipse within 4.0--10.1~$\MJ$
and 9.0--9.3~$\Sunits$. The large uncertainty in the mass (the width of the ellipse) is dominated
by the uncertainty in the gravity, since radius is roughly constant at these entropies.
We note that non-Gau{\ss}ian errorbars are trivial to propagate \Ae{through} when determining \Ae{the} mass and entropy in this way since
it is only a matter of mapping each $(\log g,\Teff)$ pair to an $(M,S)$ point.

Since this $M$--$S$ determination concerns only the current state, it is independent of the cooling sequence,
in particular of the `hot vs.\ cold start' issue.
Nevertheless, with this approach, it is immediately apparent what constraints the age imposes on the initial conditions.
Indeed, not all $(M,S)$ are consistent with an age since no planet of a given mass can be at a higher entropy than
the hot-start model at that time (see Fig.~\ref{fig:coolingS}). The entropy of hot starts is shown in Fig.~\ref{fig:MSloggT_eg}
after 10, 30, and 50~Myr. The 30-Myr age excludes objects with $M\la6~\MJ$, which is the hot-start mass of this example.
Considering only hot-start evolution sequences would have been equivalent to requiring the solution to be on one of the hot-start
(grey) curves. This is however a restriction which currently could not be justified, given our ignorance about the outcome of the formation processes.

The \Ae{solid} line in Fig.~\ref{fig:MSloggT_eg} indicate the derived constraints on the initial entropy, assuming an age of 30~Myr.
These constraints are similar to ones based on luminosity (see Fig.~\ref{fig:MSplotnoD}) but are somewhat tighter since an upper
mass limit is provided by the measurement of $\log g$. Even within a set of models, i.e.\ putting aside possible systematic
issues with the atmospheres of young objects, it is however often the case that the surface gravity is rather ill determined
(as for the objects discussed below in Section~\ref{sec:obsobj}).
In this case, provided a sufficiently large portion of the spectrum is covered,
we expect the approach based on the bolometric luminosity presented in this work to be more robust   %
than the derivation of constraints from effectively only the surface temperature. %
Indeed, the former avoids compounding uncertainties in $\Teff$ with those in the radius in an evolutionary sequence,
which can be further affected by the presence of a core (of unknown mass).
With both sustained modelling efforts and the detailed characterisation of an increasing number of detections,
one may hope that reliable atmosphere models for young objects will become available in a near future,
allowing accurate determinations of the mass, radius, and initial entropy of directly-detected exoplanets.

\section{Comparison with observed objects}
\label{sec:obsobj}

\subsection{Directly-detected objects}
\label{sec:ddo}
\citet{neuh12} recently compiled and homogenously analysed photometric and spectral data for
directly-imaged objects and candidates,
selecting only those for which a mass below $\simeq25$~$\MJ$ is possible\footnote{
This value was chosen by \citet{schneider11} as an approximation to the
`brown-dwarf desert', which is a gap in the mass spectrum between $\simeq25$--90~$\MJ$
\citep{marcy00,grether06,luhm07,dieter12}.
}.
They report \Ae{luminosities and effective temperatures}, which they either take from the
\Ae{discovery articles} or calculate, usually from bolometric corrections %
when a spectral type or colour index is available or brightness difference with the primary when not.
\citeauthor{neuh12} then use a number of hot-start cooling models
to derive (hot-start) mass values along with errorbars,
while recognising that hot starts suffer from uncertainties at early ages.

In this section, we %
determine joint constraints %
on the masses and initial entropies of directly-detected objects,
focusing on the ones for which (tentative) additional mass information is available.
A proper statistical analysis of the set of $M(\Si)$ curves
would be challenging at this point due to the inhomogeneity of the observational campaign designs.
However, upcoming surveys should produce sets of observations with well-understood and homogenous biases,
convenient for a statistical treatment.

\begin{figure}
\includegraphics[width=84mm]{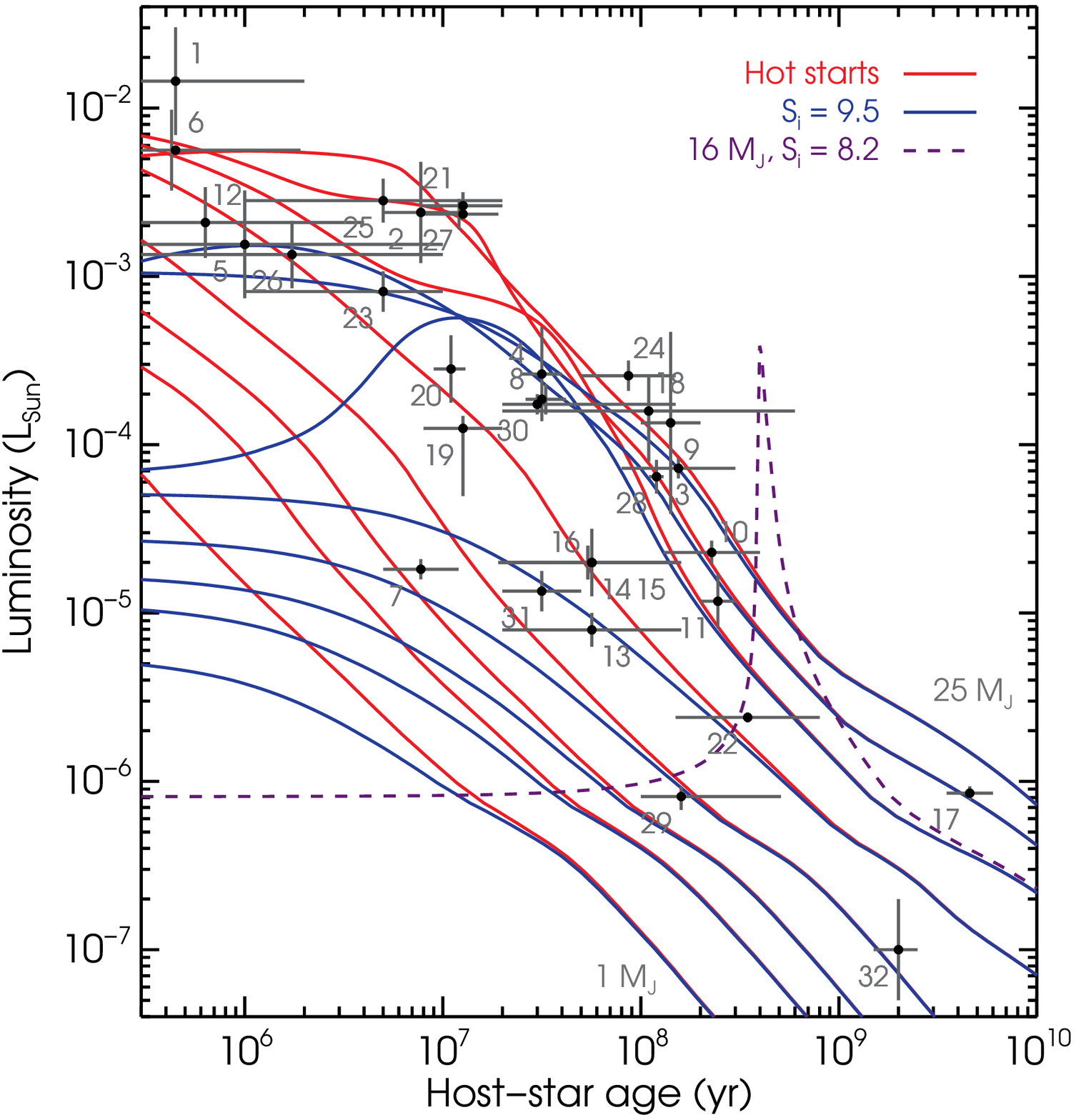}
\caption{
Directly-observed objects with a hot-start mass \Ae{below} $\sim25~\MJ$ compared with
cooling curves for $M=1, 2, 3, 5, 10, 15, 20$, and 25~$\MJ$ (bottom to top), for $\Si=9.5$ (dashed) and hot starts (dash-dotted);
the 25-$\MJ$ curve is from \citet{burr97} and has no cold-start equivalent.
\Ae{A lightcurve with a late-time deuterium `flash' (at 400~Myr for $M=16~\MJ$, $\Si=8.2$) is also shown
to draw attention that some of these objects could, in principle, be undergoing rapid deuterium burning
(Marleau \& Cumming, in prep., but see \citealp{boden13}, who independently found such flashes).}   %
The data points \Ae{are based on \citealp{neuh12} (see text for details) and} are,
with in boldface those investigated more closely below:   %
(1)~GG Tau~Bb, (2)~TWA 5~B, (3)~GJ 417~BC, (4)~GSC 8047~B/b,
(5)~DH Tau~B/b, (6)~GQ Lup~b, \textbf{(7)~2M1207~b}, (8)~AB Pic B/b, (9)~LP 261-75~B/b, (10)~HD 203030~B/b,
(11)~HN Peg~B/b, (12)~CT Cha~b, %
\boldmath{$(13,14,15,16)$}\textbf{~HR 8799~bcde},
(17)~Wolf 940~B/b, (18)~G 196-3~B/b, \textbf{(19)~\boldmath{$\beta$} Pic~b}, (20)~RXJ1609~B/b, (21)~PZ Tel~B/b,
(22)~Ross 458~C, (23)~GSC 06214~B/b, (24)~CD-35 2722~B/b, (25)~HIP 78530~B/b, (26)~SR 12~C, (27)~HR 7329~B/b, %
\Ae{(28)~2M0122~b, (29)~GJ 504~b, (30)~$\kappa$ And~b, (31)~2M0103 ABb, (32)~WD 0806-661~B/b}.
\Ae{Note that some luminosity errorbars were shifted for clarity
and that those of Ross 458~C ($0.03$~dex) are smaller than its symbol.
}
}
\label{fig:Neuhaeuser}
\end{figure}

Before turning to specific objects, we display
in Fig.~\ref{fig:Neuhaeuser} the data collected and computed by \citet{neuh12}, \Ae{as well as more recent detections,}
along with hot- and cold-start cooling tracks for different masses. This is an update of
the analogous figures of \citet{marl07} and \citet{janson11}, which had only a handful
of data points.
\Ae{Given} doubts about its nature \citep[e.g.][]{janson12,curr12_fomal,kalas13,kenw13,gali13,curr13_fomal},
we \Ae{do not include} the reported upper limit for Fomalhaut~b in this plot.
Since their uncertainties are large, the central age values are taken as the geometric mean of
the upper and lower bounds reported \Ae{if no value is given}. For RXJ1609~B/b, we use instead $11\pm2$~Myr \citep*{pecaut12}.
The errorbars for the HR~8799 planets go up to 160~Myr and do not include the \Ae{controversial} \citet{moya10b}
asteroseismology\footnote{\Ae{The linguistically inclined reader will delight in the communication of \citet{gough} about the term's prefix `ast(e)ro-'.}}
measurement of 1.1--1.6~Gyr since it is not used in our analysis (see \Ae{discussion} in Section~\ref{sec:8799Lt}).
\Ae{Finally, since no luminosity was given for WD~0806-661~B/b, we crudely estimate from the hot-start mass of 6--9~$\MJ$
from \citet{luhm12WD} and the \citet{spiegel12} models a luminosity of $\log L/\LSun=-7.0\pm0.3$ at 1.5--2.5~Gyr.}

\Ae{Fig.~\ref{fig:Neuhaeuser} also includes four recent objects discovered since the analysis of \citet{neuh12}:
2M0122~b \citep{bowler13}, GJ 504~b \citep{kuzu13}, 2M0103~ABb \citep{delorme13}, and $\kappa$~And~b \citep{carson13,bonn13_kap}.
For 2M0103~ABb, we estimate a bolometric luminosity of $\log L/\LSun = -4.87\pm0.12$ as done in Appendix~\ref{app:betaPic}
for $\beta$~Pic~b. The same approach with $\kappa$~And~b yields $\log L/\LSun = -3.83\pm0.15$, which is entirely consistent with the
published value of $-3.76\pm0.06$~dex \citep{bonn13_kap}. We show the conservative age range of $30_{-10}^{+120}$~Myr for $\kappa$~And~b.}

\Ae{A recent detection which is not included in Fig.~\ref{fig:Neuhaeuser} is a candidate companion to HD 95086
with a hot-start mass near 4~$\MJ$ \citep{rameau95086,meshkat13}, since only an $L'$-band measurement is available.
Nevertheless, we report a prediction for its luminosity of $\log \Lbol/\LSun \simeq -4.8\pm0.4$ from the estimated $\Teff =1000\pm200$~K
and $\log g = 3.9\pm0.5$~(cm\,s$^{-2}$), with the $\Lbol$ errorbars entirely dominated by those on $\Teff$
and ignoring that the atmospheric parameters were in fact estimated from hot-start models.}

Fig.~\ref{fig:Neuhaeuser} shows that there are already many data points which -- at least based solely on their luminosity -- could be explained
by cold, warm, or hot starts, highlighting the importance of being open-minded about the initial entropy
when interpreting these observations.
\Ae{Indeed, as \citet{morda12_I} carefully argue, it is presently not warranted to assume a unique mapping between core accretion (CA)
and cold starts on the one hand, and gravitational instability and hot starts on the other hand.
(Even in the case of a weak correlation, planets found beyond $\simeq50$~AU, the farthest location where
CA should be possible \citep{rafikov11}, could still in principle have formed by core accretion
and then migrated outward \citep[e.g.][]{ida13}.)
As an extreme example of a cold start, we also display a cooling curve for a deuterium-burning object with a low $\Si$ which
undergoes a `flash' at late times, somewhat arbitrarily chosen to pass near the data point of Ross 458~C \citep{burg10};
this contrasts with a monotonic hot-start cooling track at $\simeq12~\MJ$ which would also match the data point.
Such solutions will be explored in a forthcoming work but we already note that, very recently,
\citet{boden13} independently found lightcurves with flashes to be a possible outcome of a realistic formation process.}

There are two features of the data distribution in Fig.~\ref{fig:Neuhaeuser} which immediately stand out.
The first is that the faintest young objects are brighter than the faintest oldest objects,
i.e.\ that the minimum detected luminosity decreases with the age of the \Ae{companion}.
Moreover, this minimum, with the exception of
data points (7), \Ae{(31), and (30)}  %
(2M1207~b and \Ae{WD 0806-661~B/b, which are particular for different reasons, and GJ 504~b})
approximately follows the cooling track of \Ae{a} hot-start \Ae{planet} of $\simeq10~\MJ$.
\Ae{The interpretation of this fact is not obvious given that the data points forming
the lower envelope come from multiple surveys and that different observational biases apply at
different ages (e.g.\ due to the relatively low number of young objects in the solar neighbourhood).
}  %

The second feature is the absence of detections between the hot-start 10- and 15-$\MJ$ cooling curves,
roughly between 20 and 100~Myr.
More accurately, there is, in a given age bin in that range, a lower density of data points with luminosities
around 10$^{-4}~\LSun$ than at higher or lower luminosities.
A proper assesment of the statistical significance of this `gap' in the data points would require taking both
the smallness of the number of detections around 40~Myr and the biases and non-detections of the
various surveys into account.
However, it would not be surprising if the underdensity in the luminosity function ${\rm d}N/{\rm d}L$  of the data points,
at a fixed age, proved to be real,
since there is also a suggestive underdensity in the cooling curves. %
Indeed, the onset of deuterium burning near 13.6~$\MJ$ %
slows down the cooling, which breaks the hot-start scaling $L\propto t^{-1}$ (see Section~\ref{sec:gencooling}) and 
leads to a greater distance between the hot-start cooling curves for 10 and 15~$\MJ$ than for 5 and 10~$\MJ$.
(This is clearly visible in fig.~1 of \citet{burr01}, which also shows that there is a similar gap for low-mass stars
at 10$^{-4}~\LSun$ and 1--10~Gyr, due to the hydrogen main sequence.)
In particular, cooling tracks for objects of 15, 20, and 25~$\MJ$ nearly overlap at $\sim100$~Myr around 10$^{-4}~\LSun$,
where data points and their errorbars collect too. This tentative indication of an agreement between the detections
and the cooling tracks
suggests that the latter might be consistent with the data\footnote{
Cold-start curves too show this gap, to the extent that the luminosity rise due to deuterium burning
is very sensitive to the initial entropy (see Fig.~\ref{fig:Neuhaeuser}),
which would need to be set accordingly finely to have the lightcurves pass through
the data gap.
The implicit assumptions here
are that the observed distribution of masses is uniform in the approximate range 5--25~$\MJ$ (as are the mass values chosen
for Fig.~\ref{fig:Neuhaeuser}), and that the same applies to the initial entropy.
The former cannot currently be validated but the latter seems reasonable, as the entropy interval over which cooling curves
change from going above to below the gap is very narrow.
}. 
It will therefore be interesting to see how significant the `gap' is and how it evolves as
data points are added to this diagram.

\Ae{In} Fig.~\ref{fig:Neuhaeuser}, the best-fitting age is calculated as the geometric mean
of the reported upper and lower bounds since, in most cases, no best-fitting value is provided, and the bounds
are typically estimates from different methods, which cannot be easily combined.
In fact, the ages of young ($\la500$~Myr) stars are in general a challenge to determine, as \citet{soderblom10} reviews,
and represent the main uncertainty in direct observations.  %
Moreover, as \citet{fort05} point out, assuming co-evality of the companion and its primary may be
problematic for the youngest objects. Indeed, a formation time-scale of $\simeq1$--10~Myr in the core-accretion scenario
would mean that some data points of Fig.~\ref{fig:Neuhaeuser} below $\sim10$~Myr may need to be shifted to \Ae{significantly}
lower ages, by an unknown amount. This consideration is thus particularly relevant for
GG~Tau~Bb, DH~Tau~B/b, GQ~Lup~b, and CT Cha~b (data points 1, 5, 6, and 12), which are all possibly younger
than 1~Myr, and would require a closer investigation.

We now provide detailed constraints for three planetary systems,    %
chosen for the low hot-start mass of the companion (2M1207)
or because additional mass information is available (HR~8799 and $\beta$~Pic).

\subsection{2M1207}
\label{sec:MS1207}

The companion to the brown dwarf 2MASSWJ 1207334--393254 (2M1207~A, also known as TWA~27~A; \citealp{gizis02}) is
the first directly-imaged object with a hot-start planetary mass \citep{chauv04,chauv05}. 
Since the age and luminosity of 2M107~b are the inputs for our analysis, they are
discussed in some detail in Section~\ref{app:2M1207}, along with tentative information on the mass.
We adopt an age of $8^{+4}_{-3}$~Myr \citep{chauv04,song06}
and a luminosity of $\log L/\LSun =-4.68\pm0.05$ \citep{barman11_1207},
and assume that deuterium-burning masses above $\simeq13$~$\MJ$ \citep{spiegburrmils11}   %
are excluded.

\begin{figure}
\includegraphics[width=84mm]{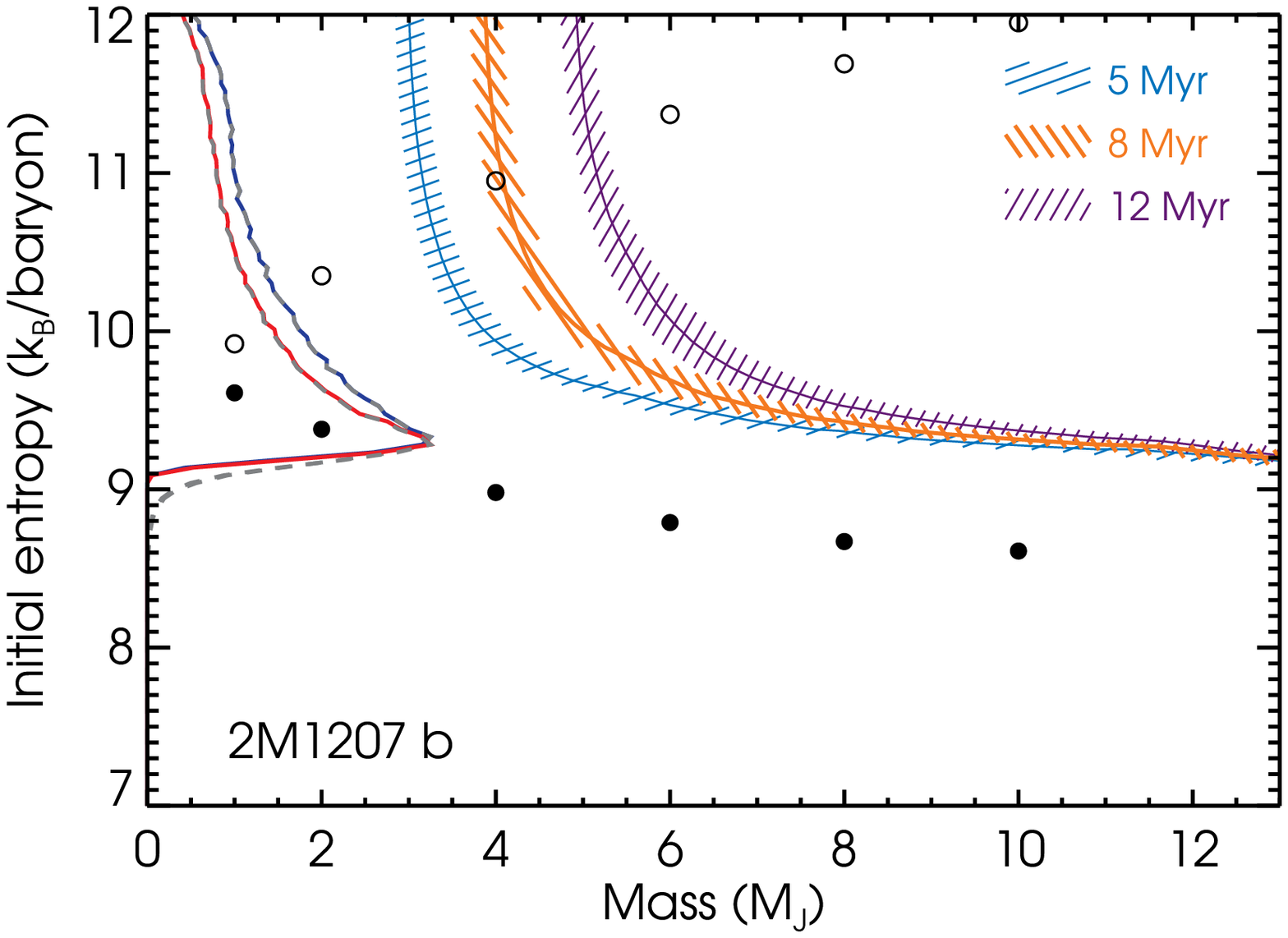}
\caption{
Allowed values of planet mass $M$ and initial entropy $\Si$ for 2M1207~b for assumed ages of
$8_{-3}^{+4}$~Myr.  %
The luminosity is that of \citet{barman11_1207}, $\log \Lbol/\LSun=-4.68\pm0.05$,
with the hatched regions corresponding to curves within $\pm1~\sigma$ of the luminosity values and
the thick lines to the central value.
The curves along the vertical axis show posterior distributions from MCMCs with different mass priors
and assumed luminosity distributions. The solid curves only have an upper mass limit of 13~$\MJ$,
while the dashed curves include solutions in the deuterium-burning region, up to 20~$\MJ$.
For the blue curve (further from the left axis) and the corresponding dashed grey posterior (practically identical),
a lognormal error distribution in time centred at $\sqrt{5\times12}\simeq7.7$~Myr and of width 0.19~dex was used,
while the red curve and corresponding dashed line are for a flat distribution in time between 5 and 12~Myr (and zero otherwise).
The circles show the results from \citet{marl07}, increased by 0.38~$\Sunits$ (cf.\ Fig.~\ref{fig:coolingcurvesSpiegel}),
for cold starts (filled circles) and hot starts (open circles), where the entropy is 1~Myr after the onset of cooling.
}
\label{fig:MSplot_1207}
\end{figure}

Fig.~\ref{fig:MSplot_1207} shows the joint constraints on the mass and initial entropy of 2M1207~b
based on its luminosity and age. We recover the hot-start mass of 3--5~$\MJ$ \citep{barman11_1207},
\Ae{with equation (\ref{eq:Mhs}) predicting $\simeq3.9~\MJ$,}
but also find solutions at higher masses.
If deuterium-burning masses can be excluded, the formation of 2M1207~b must have led
to an initial entropy of $\Si\geqslant9.2$, with an approximate formal uncertainty on this lower bound
of 0.04~$\Sunits$ (see Section~\ref{sec:MSishape}) due solely to the luminosity's statistical error,
independent of the age's.  %
This initial entropy implies that the \citetalias{marl07} cold starts are too cold by 0.7~$\Sunits$,
roughly independently of the mass, to explain the formation of this planet.
This is consistent with the time-scale-based conclusion of \citet*{lodato05}
that core accretion cannot be responsible for the formation of this system {\em if} one also accepts
the received wisdom that core accretion necessarily leads to the coldest starts.
However, our robust quantitative finding is more general,
in that it provides constraints on the result of the formation process which are {\em model-independent}.

To show how these $M(\Si)$ constraints can easily be made even more quantitative and thus suitable for
statistical analyses, we ran a Metropolis--Hastings Markov-chain Monte Carlo (MCMC; e.g.\ \citealp{greg05})
in mass and entropy with constant priors on these quantities. %
Uncertainties in the luminosity $L$ and age $t$ were included in the calculation of the $\chi^2$
by randomly choosing an `observed' $L_{\rm obs}$ and a stopping time for the cooling curve $t_{\rm stop}$
at every step in the chain.
The quantity $\log L_{\rm obs}$ was drawn from a Gau{\ss}ian defined by the reported best value and its errorbars,
and $t_{\rm stop}$ from a distribution which is either
constant in $t$ between the adopted upper and lower limits $t_1=5$~Myr and $t_2=12$~Myr and zero otherwise,
or lognormal in $t$, centred at $t_0 = \sqrt{t_1t_2} = 7.7$~Myr and
with $\sigma_{\log t} =\log t_2/t_1=0.19$~dex.
The results are shown along the vertical axis of Fig.~\ref{fig:MSplot_1207} for four different assumptions.
To obtain the two solid lines, we applied an upper mass cut at 13~$\MJ$ and took a lognormal
(less peaked curve) or a top-hat (more peaked) distribution in time.
The dashed curves come from the same MCMC chains but with a mass cut-off of 20~$\MJ$,
i.e.\ including deuterium-burning objects. Because of the delayed cooling due to deuterium burning,
the required initial entropy drops down faster with mass than in the cold-start branch,
such that lower $\Si$ values are possible. At 20~$\MJ$, the required $\Si$ is 8.2,
but it is still only 9~$\Sunits$ at 15~$\MJ$. However, the phase space
for high masses is very small since the initial entropy needs to be extremely finely tuned;
hence the smallness of the effect on the posteriors.
As Fig.~\ref{fig:MSplot_1207} shows, different assumptions on the luminosity, age, and mass priors
all lead to similar results for the initial entropy, namely that $\Si\geqslant9.2$ and
that there are more solutions near this lower limit.

\subsection{HR~8799}
\label{sec:8799}

We now turn to the only directly-imaged system with multiple objects for which planetary masses are possible, HR~8799.
The age of the system and the luminosities of the companions are 
discussed in Section~\ref{app:8799}, along with information on the mass.
We consider ages of 20 to 160~Myr, close to the values of \citet{marois08}, and use the standard
luminosities of $\log \Lbol/\LSun=-5.1 \pm 0.1$ (HR~8799~b),
$-4.7 \pm 0.1$ (cd) and $-4.7 \pm 0.2$ (e) from \citet{marois08,marois10}.
It also seems reasonable to assume that deuterium-burning masses can be excluded for all objects,
thanks to the (preliminary) results from simulations of the system's dynamical stability.

\begin{figure}
\includegraphics[width=84mm]{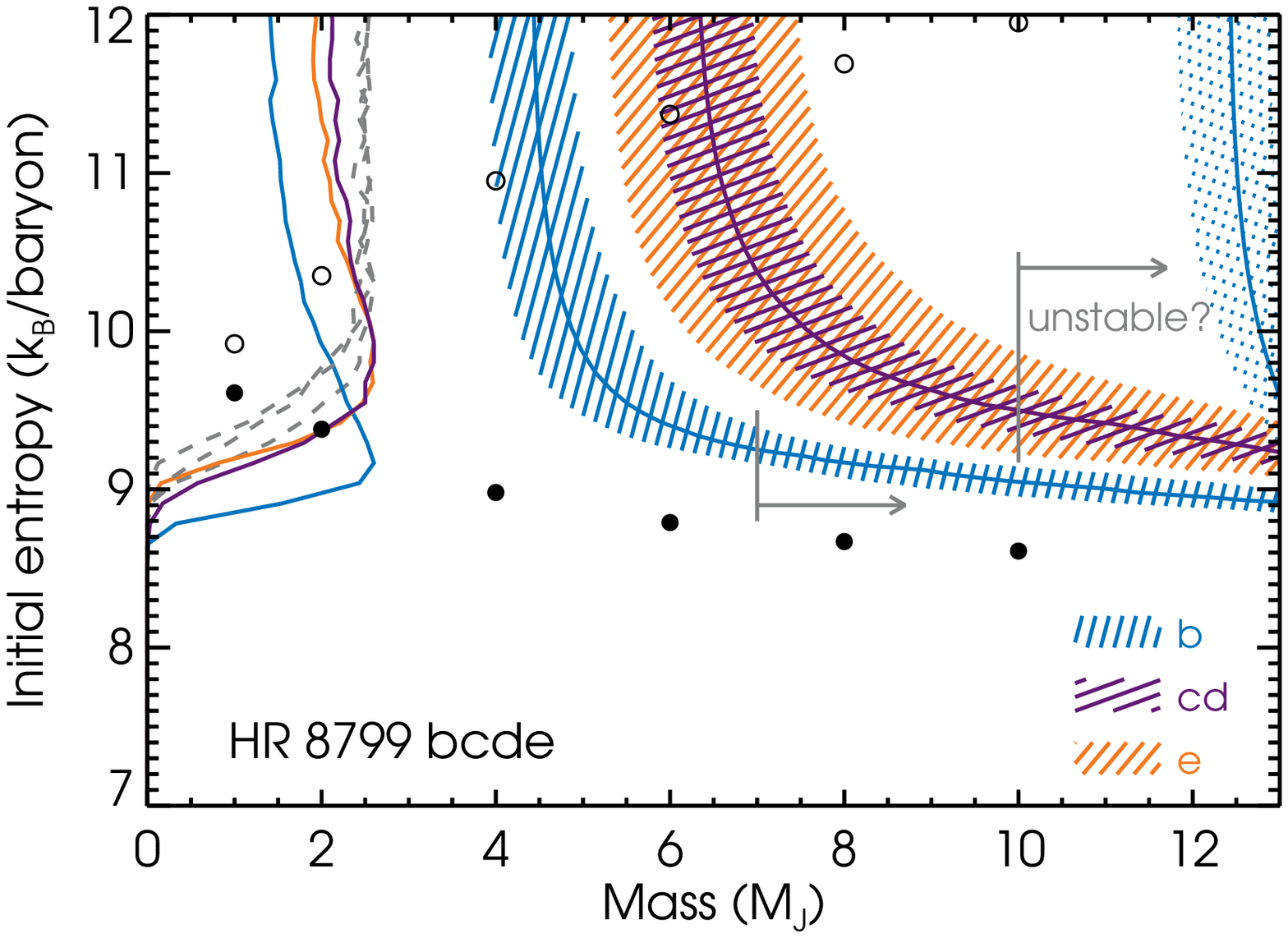}
\caption{
Allowed values of planet mass $M$ and initial entropy $\Si$ for HR~8799~b ($\log \Lbol/\LSun=-5.1 \pm 0.1$; blue bands),
HR~8799~cd ($-4.7 \pm 0.1$; purple), and HR~8799e ($-4.7 \pm 0.2$; orange) for assumed ages of 20 (lower-mass group) %
and 160~Myr (higher-mass band; planets cde are outside the plotting range).
Luminosities are from \citet{marois08,marois10},  %
and nearly in agreement with \citet{marl12}.
The lines correspond to the best-fitting luminosity values, while the hatched regions use the 1-$\sigma$ errorbars.
Along the left axis are shown the posterior distributions in $\Si$ (colours as for the $M$--$\Si$ hashed bands),
with no particular relative normalisation.  %
These were obtained from an MCMC, taking luminosity and age uncertainties into account based on lognormal distributions
(see text), and using flat priors in $\Si$ and $M$.
The mass upper cut-off is $M_{\rm max}=13~\MJ$ (full lines) or $M_{\rm max}=(7,10,10,10)~\MJ$
(for planets bcde, respectively; dashed lines), since analyses of the system's dynamical stability
seem to indicate that higher masses are unstable.
This is indicated by the vertical line segments with arrows in the direction of the excluded masses.
The circles show the results from \citetalias{marl07}, increased by 0.38~$\Sunits$ (cf.\ Fig.~\ref{fig:coolingcurvesSpiegel}),
for cold starts (filled circles) and hot starts (open circles), where the entropy is 1~Myr after the onset of cooling.
}
\label{fig:MSplot_8799}
\end{figure}

Fig.~\ref{fig:MSplot_8799} shows the joint constraints on the masses and initial entropies of
HR~8799~b, d, and e. Uncertainties in the age are taken into account by considering the two extremes
of 20~Myr and 160~Myr separately, while the 1-$\sigma$ errors in the luminosities are reflected by the width of the hatched regions.
We find hot-start masses for 20~Myr of $4.4_{-0.5}^{+0.4}$~(b), $6.3\pm0.6$~(cd), and $6\pm1~\MJ$~(e),  %
where the errorbars here come only from those on the luminosity,
\Ae{fully consistent with the prediction by equation (\ref{eq:Mhs}) of $\simeq4.2$ or 6~$\MJ$.}
These values are in good agreement with \citet{neuh12} and are similar to the
usually-cited 30-Myr values of $(5, 7, 7, 7)~\MJ$ \citep{marois10}. %
The hot-start masses for 160~Myr are above 12 (HR~8799~b) and 13~$\MJ$ (cde),
and the $M(\Si)$ constraints for the latter three are not shown since they are within the deuterium-burning regime.

Excluding deuterium-burning masses for all objects and using only the luminosity measurements,
Fig.~\ref{fig:MSplot_8799} shows that all planets of the HR~8799 system must have formed with an initial entropy
greater than 9~$\Sunits$, with $\Si\ga8.9$ for~b, $\Si\ga9.2$ for~c and~d, and $\Si\ga9.1$ for~e
\Ae{(using as throughout this work the published \citetalias{scvh} entropy table; see Appendix~\ref{sec:Sdiff}).}
Using tentative upper mass limits of 7 and 10~$\MJ$, respectively, the lower bounds on the initial entropies
can be raised to 9.2 (b) and 9.3~$\Sunits$ (cde) if one takes the conservative scenario of the 1-$\sigma$
lower luminosities value at 20~Myr. These lower bounds on the entropy are however mostly independent
of the age because they are set by cold-start solutions, where the age is much smaller than the cooling time
at that entropy. Formal uncertainties on the lower bounds due to those in the luminosities are (see Section~\ref{sec:MSishape})
approximately 0.07 (bcd) or 0.14~$\Sunits$ (e) and thus negligible.

Here too we ran an MCMC to derive quantitative constraints on the initial entropy of each planet.
The quantity $\log L_{\rm obs}$ was drawn from a Gau{\ss}ian defined by the reported best value and its errorbars,
and $t_{\rm stop}$ from a distribution which is either
constant in $t$ between the adopted upper and lower limits $t_1=20$~Myr and $t_2=160$~Myr and zero otherwise,
or lognormal in $t$, centred at $t_0 = \sqrt{t_1t_2} = 57$~Myr and
with $\sigma_{\log t} =\log t_2/t_1=0.9$~dex.
Posteriors on the initial entropy for each of the HR~8799 planets are shown along the vertical axis
in Fig.~\ref{fig:MSplot_8799}, using a flat prior in $\Si$ and a mass prior constant up to an $M_{\rm max}$
and zero afterward.    %
The cases `without mass constraints' ($M_{\rm max}=13~\MJ$) are nearly constant in $\Si$, especially for planet~b,
but show a peak near cold-branch values of 9 and 9.5~$\Sunits$ for HR~8799~b and~cde.
Adding mass information from dynamical-stability simulations by taking $M_{\rm max} = (7,10,10,10)~\MJ$
flattens the $\Si$ posterior and shifts the minimum bounds at half-maximum
from $(8.9, 9.2, 9.2, 9.2)$ to $(9.3, 9.5, 9.5,9.4)~\Sunits$, respectively.
We note that these results are insensitive to both the form of the uncertainty in $t$
and the use of a non-flat prior in mass (as shown below for $\beta$~Pic~b in Section~\ref{sec:MSbetaPicSi}).

Comparing to the `tuning fork' entropy values reproduced in Fig.~\ref{fig:MSplot_8799},
we find that the coldest-start models of \citetalias{marl07} cannot explain the luminosity measurements
for the HR~8799 planets. \citet{spiegel12} and \citet{marl12}
also came to this conclusion, with the latter noting that `warm starts'  %
match the luminosity constraints. It is now possible to say specifically that {\em the \citet{marl07} cold starts
would need to be made $\Delta S\simeq0.5~\Sunitsem$ hotter to explain the formation of the HR~8799 planets}.
Given that the precise outcomes of the core accretion and gravitational instability scenarios are uncertain
and that this system represents a challenge for both (as \citealp{marois10} and \citealp{curr11} review),
quantitative comparisons such as our procedure allows should be welcome to help evaluate the plausibility
of the one or the other. %

We note in passing that one needs to take care also when interpreting the measurements of
\citet{hink11_8799} and \citet{close10}. These authors measured upper limits on the brightness of
companions within 10~AU and between 200 and 600~AU from the star, respectively. However, both groups then used
the hot-start models of \citet{bara03} to translate the brightness limits into masses
(11~$\MJ$ at 3--10~AU and 3~$\MJ$ within 600~AU, \Ae{respectively}).
Therefore, since colder-start companions would need to be more massive to have the same luminosity,
what they provide are really \Ae{``}\emph{lower} upper limits\Ae{''} on the mass of possible companions.
How much higher the masses could realistically be in this case is difficult to estimate
without a bolometric luminosity, but there is an important general point:
without the restriction of considering only hot-start evolutionary tracks,   %
{\em luminosity upper limits do not provide unambiguous mass constraints}.
Incidentally, this more general view of the results of \citet{hink11_8799}  %
means that the unseen companion evoked by \citet{su09} as the possible cause of the inner hole (at $\la6$~AU)
does not have to be of small mass. \Ae{However,} this inner object would nevertheless have to be consistent with the results of
dynamical stability simulations, \Ae{with those of \citet{gozmiga13} indicating a mass less than $\simeq1$--8~$\MJ$}.

\subsection[beta Pic]{\ensuremath{\bmath{\beta}}~Pic}
\label{sec:MSbetaPic}
\label{sec:MSbetaPicSi}

\begin{figure}
\includegraphics[width=84mm]{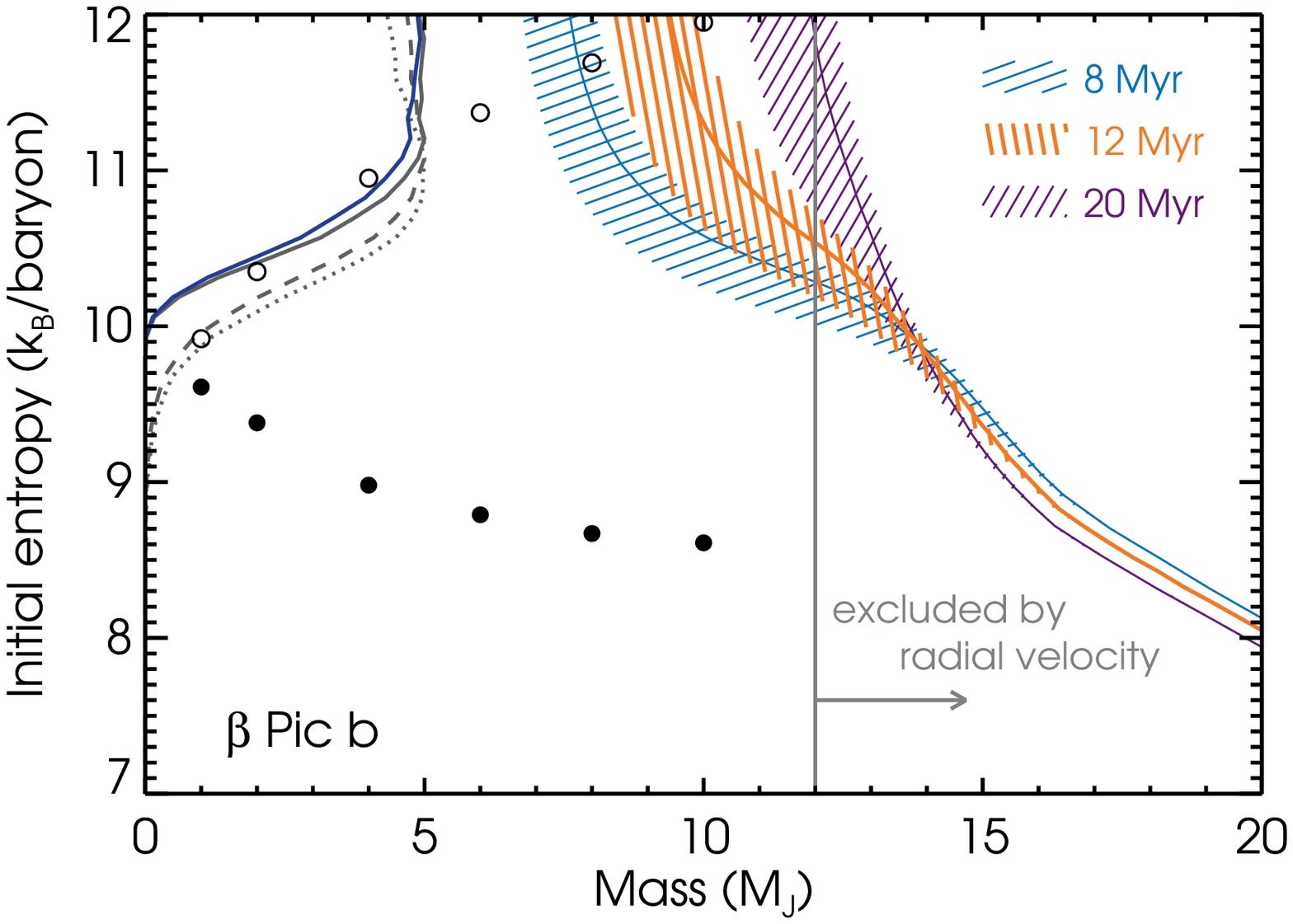}
\caption{
Allowed values of planet mass $M$ and initial entropy $\Si$ for $\beta$~Pic~b using
our estimated $\log\Lbol/\LSun=-3.90^{+0.05}_{-0.12}$
and an age of $12^{+8}_{-4}$~Myr. %
The hatched regions correspond to luminosity values within the (asymmetrical) 1-$\sigma$ interval, with the central values
marked by thick lines.
The curves along the vertical show the marginalised posterior distribution on $\Si$ from an MCMC simulation
using the luminosity and age values and their uncertainties. These were taken to be lognormal but asymmetric for the luminosity,
and lognormal in time, centred at $t_0 = \sqrt{8\times20} = 12.7$~Myr with a standard deviation of 0.35~dex.
The full blue line (closer to the vertical axis) results from the $(M,\Si)$ distribution being multiplied
by a ${\rm d}N/{\rm d}M \propto M^{\alpha}$ prior,
with $\alpha = -1.3$ \citep{ac08,nielsen10}, and zero for $M>12~\MJ$,
while the full grey posterior has only a mass cut at 12~$\MJ$, from the \citet{lagr12_rv} radial-velocity constraint.
The dashed and dotted curves both have an upper mass cut of 20~$\MJ$, with and without the ${\rm d}N/{\rm d}M$ prior applied,
respectively.
The circles show the results from \citet{marl07}, increased by 0.38~$\Sunits$ (cf.\ Fig.~\ref{fig:coolingcurvesSpiegel}),
for cold starts (filled circles) and hot starts (open circles), where the entropy is 1~Myr after the onset of cooling.
}
\label{fig:MSplot_betaPic}
\end{figure}

A companion to the well-studied star $\beta$~Pic was first observed in 2009 \citep{lagr09,bonn11}
and, very recently, it became the first directly-detected object with a planetary mass
for which radial-velocity data are also available. The age of the system is \Ae{taken as} $12^{+8}_{-4}$~Myr \citep{zucker01},
and we discuss in detail in Section~\ref{app:betaPic} 
how we derive a bolometric luminosity\footnote{
As this manuscript was being prepared, we became aware of the first robust estimate
of the bolometric luminosity, by \citet{bonn13}. They find
$\log \Lbol/\LSun = -3.87\pm0.08$, which excellently agrees with our value and thus does not change our conclusions.
In particular, they find similar constraints on the initial entropy of $\beta$~Pic~b,
although this depends on which band they use
(cf.\ their fig.~11 with our Fig.~\ref{fig:MSplot_betaPic}).
\Ae{We also note the more recent estimate by \citet{curr13} of $\log \Lbol/\LSun = -3.80\pm0.02$
(very near our approximate 1-$\sigma$ upper limit),
from which they estimate a (hot-start) mass in the 3--11-$\MJ$ range.  %
Importantly, they obtain the lower masses by considering an age of 7~Myr for $\beta$~Pic~b,
i.e.\ by relaxing the assumption that it is co-eval with its star (see Section~\ref{sec:ddo}).}
}  of $\log \Lbol/\LSun = -3.90^{+0.05}_{-0.12}$.

Fig.~\ref{fig:MSplot_betaPic} shows the $M(\Si)$ constraints available for $\beta$~Pic~b
from our luminosity estimate and the radial-velocity (RV) constraint.
We recover a hot-start mass $\simeq9.5\pm2.5~\MJ$ \Ae{(cf.\ $\simeq9.4~\MJ$ from equation \ref{eq:Mhs})},
in agreement with \citet{quanz10} and \citet{neuh12},
but additionally find that higher masses are consistent with the luminosity measurement.
Excluding solutions where deuterium burning plays an important role in the evolution of the object
(recognisable by the extreme thinness of the constant-luminosity $M(\Si)$ curve) implies that
$\Si\geqslant9.8$.  %
Using the RV mass upper limit, these constraints on the initial entropy can be made tighter:
with an age of 12~Myr, it must be that $\Si\geqslant10.5$.
Since this corresponds to a warm start, both uncertainties on the age and on the luminosity
contribute to that on the minimum $\Si$, of order 0.5~$\Sunits$.

This lower limit on $\Si$ implies that coldest-start objects of any mass are too cold by a significant 1.5--2.0~$\Sunits$.
Various authors \citep[e.g.][]{quanz10,bonn11} recognised that the classical cold starts (\citetalias{marl07}; \citealp{fort08})
cannot explain the observations, but it is now possible to quantify this.
These results are mostly insensitive to the uncertainty on the age range;  %
using instead 12--22~Myr as summarised by \citet{fern08} would not change the conclusions.

As for the objects in the 2M1207 and HR~8799 systems, we ran an MCMC to obtain a posterior distribution on the initial entropy.
This is shown along the vertical axis of Fig.~\ref{fig:MSplot_betaPic} for four different assumptions.
In all cases, we assumed lognormal uncertainties on the age and the luminosity,
with asymmetric upper and lower errorbars for the latter.
For the full curve closer to the vertical axis (in blue), we applied an upper mass cut at 12~$\MJ$
and took into account that the underlying (real) mass distribution is possibly biased towards lower masses,
as radial velocity measurements indicate \citep{ac08,nielsen10}.
Out of simplicity, this was done by taking the $(M,\Si)$ distribution obtained with a flat prior in mass
and \Ae{using importance sampling to put in a posteriori} a ${\rm d}N/{\rm d}M \propto M^{\alpha}$ prior, with $\alpha = -1.3$,
thus weighing lower masses more\footnote{\Ae{\citet{wahhaj13} derive in a recent analysis of the
NICI campaign results for debris-disc stars a similar slope: if the linear semi-major axis power-law index
$\beta_a=-0.61$ as \citet{ac08} found for radial-velocity planets within 3~AU,
the 66 non-detections combined with the \citet{vigan12} survey imply $\alpha < -1.7$ to 2~$\sigma$, with the most likely values
at $0.3\alpha+\beta_a\ll-1$. When however $\beta$~Pic~b and HR~8799~bcd are included in the analysis,
$\alpha > 2.2$ for $\beta_a=-0.61$, with more solutions at $\beta_a=-2.1$, $\alpha\geqslant2.2$.  %
(Too few detected objects prevent \citet{biller13} from inferring constraints on $\alpha$ and $\beta$ in a similar analysis
of young moving-group stars.)
Note finally that %
hot-start models were used to convert magnitudes to masses
and that most targets are less than 100~Myr old, with a significant fraction near 10~Myr;
ignoring cold starts at these ages can skew the inferred (limits on the) mass distribution.  %
Also considering colder starts should yield more negative constraints on $\alpha$ given the same luminosity constraints.
}}.   %
Of course, the value of $\alpha$ might depend on the formation mechanism relevant to the object
but this serves to illustrate the effects of a non-constant prior on mass.
The other three posterior distributions on $\Si$ of Fig.~\ref{fig:MSplot_betaPic} (in grey)
correspond to the remaining combinations of `with mass cut or not'
and `with power-law mass prior or not'.
These curves are all similar,
with the radial-velocity measurement increasing the minimum bound at half maximum from
10.2 to 10.5~$\Sunits$, quite insensitively to the use of the ${\rm d}N/{\rm d}M$ prior.

At the distance from its primary where $\beta$~Pic~b is currently located ($\simeq9$~AU),
core accretion is expected to be efficient and thus a likely mechanism for its formation \citep{lagr11,bonn13}.
Thus, the question posed to formation models is whether core accretion can be made hotter (by 1.5--2~$\Sunits$)
than what traditional cold starts predict.
\Ae{Very recently, \citet{boden13} and \citet{morda13}  %
showed that in the framework of formation models (which seek to predict $\Si$), different rocky core masses
are associated with a significantly different initial entropies
at a fixed total mass; for instance, \citet{boden13} found $\Si=7.5$ for a 12~$\MJ$ object with a core of 5~$\ME$ but
$\Si=9.1$ when a different choice of parameters lead to a core mass of 31~$\ME$.}
Since these coldest starts assume that all the accretion energy
is radiated away at the shock, the constraints on the initial entropy stress the need to investigate the physics
of the shock \Ae{(and its dependence on physical quantities such as the accretion rate)},
when the initial energy content of the planet is claimed to be set.

\section{Summary}

\label{sec:discussion}

The entropy of a gas giant planet immediately following its formation is a key parameter
that can be used to \Ae{help} distinguish planet formation models \citep{marl07}.
In this paper, we have explored the constraints on the initial entropy that can be obtained for directly-detected exoplanets with a measured bolometric luminosity and age.  When the initial entropy is assumed to be very large, a `hot start' evolution, the measured luminosity and age translate into a constraint on the planet mass. In contrast, when a range of initial entropies are considered (`cold starts' or `warm starts'), the hot-start mass is in fact only a lower limit on the planet mass: larger-mass planets with lower initial entropies can also reproduce a given observed luminosity and age. Fig.~\ref{fig:MSplotnoD} shows the allowed values of mass and entropy for different ages and luminosities, and can be used to quickly obtain estimates of mass and initial entropy for any given system.

To derive these constraints, we constructed a grid of gas giant models as a function of mass and \Ae{internal} entropy which can then be stepped through to calculate the time evolution of a given planet. In a hot-start evolution, the structure and luminosity of cooling gas giant planets are usually thought of as being a function of mass and time only; \Ae{this leads to a `hot-start mass' as given by equation (\ref{eq:Mhs}).} Once the assumption of hot initial conditions is removed, however, a more convenient variable is the entropy of the planet. One way to think of this is that gas giants obey a Vogt--Russell theorem in which the internal structure, luminosity, and radius of a planet depend only on its mass and entropy \Ae{(as well as its composition, as for stars)}. Fig.~\ref{fig:SL} shows the luminosity as a function of entropy for different masses, and a general fitting formula for $L(M,S)$ is given \Ae{by equation (\ref{eq:LMSfit})}. \Ae{(Similarly, cooling tracks with arbitrary initial entropy are well described analytically by equations (\ref{eq:cool}--\ref{eq:hs_BL93}).)} A noteworthy result is that in the intermediate-entropy regime ($S\simeq 8.5$--$10~\Sunits$), where the outer radiative zone is thick and follows a radiative-zero solution, the luminosity obeys $L =Mf(S)$ as found for irradiated gas giants by \citet{ab06}, with $f(S)$ a steeply increasing function of the entropy. \Ae{We also note that constraints obtained for models with a particular helium mass fraction $Y$ can be easily translated to another $Y$ with equation (\ref{eq:S'S}) as it provides an approximate value for ${\rm d}S/{\rm d}Y$ at constant $L$. This is general and independent of the approach used to compute the cooling.}

We find that our models are in good agreement (within tens of per cent) with the hot-start models of \citet{burr97} and \citet{bara03}, as well as the cold-start models of \citet{marl07}, and cooling models calculated with the \textsc{mesa} stellar evolution code \citep{paxton11,paxton13}. \Ae{We caution that the \citet{spiegel12} and \citet{moll12} models, for example, use a version of the \citet{scvh} equation of state whose entropy is offset by a constant $0.52~\Sunits$ from the published tables (which the present work uses); this difference is not significant physically but needs to be taken into account when comparing results of various groups. Details and $(Y,P,T,\varrho,S)$ points for a quick comparison are provided in Appendix~\ref{sec:Sdiff}. The remaining intrinsic difference in entropy between our models and those just cited is then approximately $|\Delta S| \simeq 0.15$ at worse. This is more important than differences in opacities or composition}; for example, we estimate from \citet{saumon96} that the uncertainty in the helium ($Y$) and metal ($Z$) mass fractions introduces variations of at most 10 per cent in the luminosity \Ae{at a given age}.

We stress again that when the initial entropy is allowed to take a range of values, the hot-start mass \Ae{(equation (\ref{eq:Mhs}))} is only a lower limit on the companion mass. The larger range of allowed masses means that the hot-start mass could actually lead to the mischaracterization of an object, with a hot-start mass in the planetary regime actually corresponding to an object with a mass above the deuterium-burning limit for low enough entropies. One way to break the degeneracy between mass and entropy is an accurate determination of the radius from spectral fitting (or actually a determination of $\log g$ and $\Teff$ from the spectrum), which would yield the mass and (current) entropy of the object without any degeneracy (see e.g.\ Fig.~\ref{fig:MSloggT_eg}). As discussed in Section~\ref{sec:gTconstr}, however, current atmosphere models have significant uncertainties that make this approach difficult. Another possibility is to obtain independent constraints on the mass of a companion, for example from dynamical considerations.

In Section~\ref{sec:obsobj}, we applied our models to three directly-imaged objects which have hot-start masses in the planetary-mass regime and for some of which there are additional constraints on the planet mass. We find that the initial entropy of 2M1207~b is at least 9.2~$\Sunits$, assuming that it does not burn deuterium. For the planets of the HR~8799 system, we infer that they must have formed with $\Si > 9.2~\Sunits$, independent of the age uncertainties for the star.
Finally, a similar analysis for $\beta$~Pic~b reveals that it must have formed with $\Si >10.5~\Sunits$, using the radial-velocity mass upper limit of 12~$\MJ$. These initial entropy values are respectively ca.\ 0.7, 0.5, and 1.5~$\Sunits$ higher than the ones obtained from core accretion models by \citet{marl07}. This quantitatively rules out the coldest starts for these objects and constrains warm starts, especially for $\beta$~Pic~b.

An important point is that the uncertainties in age and luminosity impact the derived hot-start mass and the lower bound on initial entropy in different ways. The major uncertainty in direct detections is the age of the star and, relevant for very young systems, all the more that of the planet. This age uncertainty translates into errorbars for the hot-start mass which are $\Delta M/M\simeq \frac{1}{2}\Delta t/t$. However, the \Ae{uncertainty on the} initial entropy on the mass-independent branch of the $M(\Si)$ curve is due only to that in the bolometric luminosity, with $\Delta \Si \simeq 1/\lambda\Delta \log\Lbol$ where $\lambda\simeq0.7$ when $\Si\la 9.6$ \Ae{(or, less accurately, when $L\sim10^{-6}$--$10^{-4}~\LSun$)}. This uncertainty $\Delta \Si$ is typically very small, which means that, up to systematic errors, the initial entropy can be determined quite accurately.

It has been pointed out before that current direct-imaging detections are all inconsistent with the cold-start predictions from core-accretion models. It is important to stress however that the cold starts are in some sense an extreme case, as they assume complete radiation efficiency at the shock during runaway accretion, which is argued to set the low initial entropy of planet. By varying the nebula temperature or the accretion rate, \citet{marl07} were able to change the entropy by barely $\Delta S \simeq 0.1$; \Ae{however,} \citet{morda13} and \citet{boden13} report that it is possible within the core-accretion scenario to obtain considerably higher entropies, increased by as much as 1--2~$\Sunits$, \Ae{when considering different masses for the solid core (through a self-amplifying process explained in \citealp{morda13})}. With $\beta$~Pic~b a likely candidate for formation by core accretion, this indicates that it is essential to gain a deeper understanding about what sets the initial entropy, for instance by looking in more detail at the properties of the shock during runaway accretion.

The derived bounds on $\Si$ for the HR~8799 objects and $\beta$~Pic~b made use of information on the mass, which comes from dynamical stability analyses and radial velocity, respectively. Radial-velocity data of directly-detected planets are currently available only for $\beta$~Pic~b, but this should change in a near future as close-in planets start being detected directly. In the absence of dynamical information, an upper limit to the mass (and thus a lower limit on the initial entropy) should be obtainable from $\log g$, even if its errorbars are large; thus, in practice, the $M(\Si)$ constraint curve does not extend to arbitrarily high masses as a pure luminosity measurement would imply.

Finally, we ran Markov-chain Monte Carlo simulations to derive more detailed quantitative constraints on the mass and entropy of directly-detected objects. When taking the uncertainties in the age and luminosity into account,
we considered normal, lognormal, and flat distributions and found the chosen form to make little difference. The advantage of this approach is that it allows one to derive posterior distributions on the initial entropy, which are suitable for statistical comparisons.
Given the small semi-major axis of $\beta$~Pic~b (9~AU), we also tried a prior ${\rm d}N/{\rm d}M\propto M^{-1.3}$ (which describes the population of radial-velocity planets) in addition to a flat prior on $M$.
The latter case yielded posterior distributions on $\Si$ with a more pronounced peak.

The benefits that the expected large increase in the number of directly-detected exoplanets in the near future should bring are at least twofold. Firstly, each new detection will yield a new constraint on the initial entropy and therefore formation mechanism. Particularly with the ability to detect lower mass gas giants at small semi-major axes ($\la20$~AU), which should be possible with instruments such as GPI or SPHERE, there will be an opportunity to constrain the state of the gas giant immediately following core accretion. Secondly, with a larger sample of objects comes a chance for a statistical comparison with formation models. For example, as noted in Section~\ref{sec:ddo}, the onset of deuterium burning leads to a relative underdensity of planets with luminosities $\sim 10^{-4}~\LSun$ and ages of tens of Myr (assuming that the masses of substellar objects are smoothly distributed near the deuterium-burning limit). Indeed, it is interesting that there appears to be such an underdensity in the current data sample (see Fig. 8), although the small number of detections so far means that this could be due to a statistical fluctuation.
The best constraints on initial conditions for planet cooling will come from improved spectral models
that can give reliable determinations of $\log g$ and $\Teff$.

\section*{Acknowledgments}
We thank P.\ Bodenheimer, D.\ Saumon, J.\ Ferguson, K.\ Go{\'z}dziewski, A.\ Burrows, J.\ Fortney,
A.\ Showman, M.\ Marley, X.\ Huang, C.\ Mordasini, D.\ Spiegel, T.\ Guillot, T.\ Schmidt,
M.\ Bonnefoy, B.\ Biller, and J.\ Carson for useful and often detailed discussions,
helpful and rapid answers to inquiries, and generosity with data.
This work was supported by the National Sciences and Engineering Research Council of Canada (NSERC) and
the Canadian Institute for Advanced Research (CIFAR), and by a scholarship from the Fonds de
recherche du Qu\'ebec -- Nature et technologies (FRQNT). GDM warmly thanks the MPIA for support
during the last stages of this work.

\footnotesize{

\bibliography{paper}

}

\appendix

\section{Radii}
\label{app:R}
\Ae{
For completeness, we present and compare the radii $R$ in our models as a function of mass $M$ and entropy $S$.
This comparison thus separates out possible differences in the treatment of deuterium burning since
the planetary structure at a given $(M,S)$ is independent of the nuclear energy generation,
which only affects the time evolution.}

\begin{figure}
\includegraphics[width=84mm]{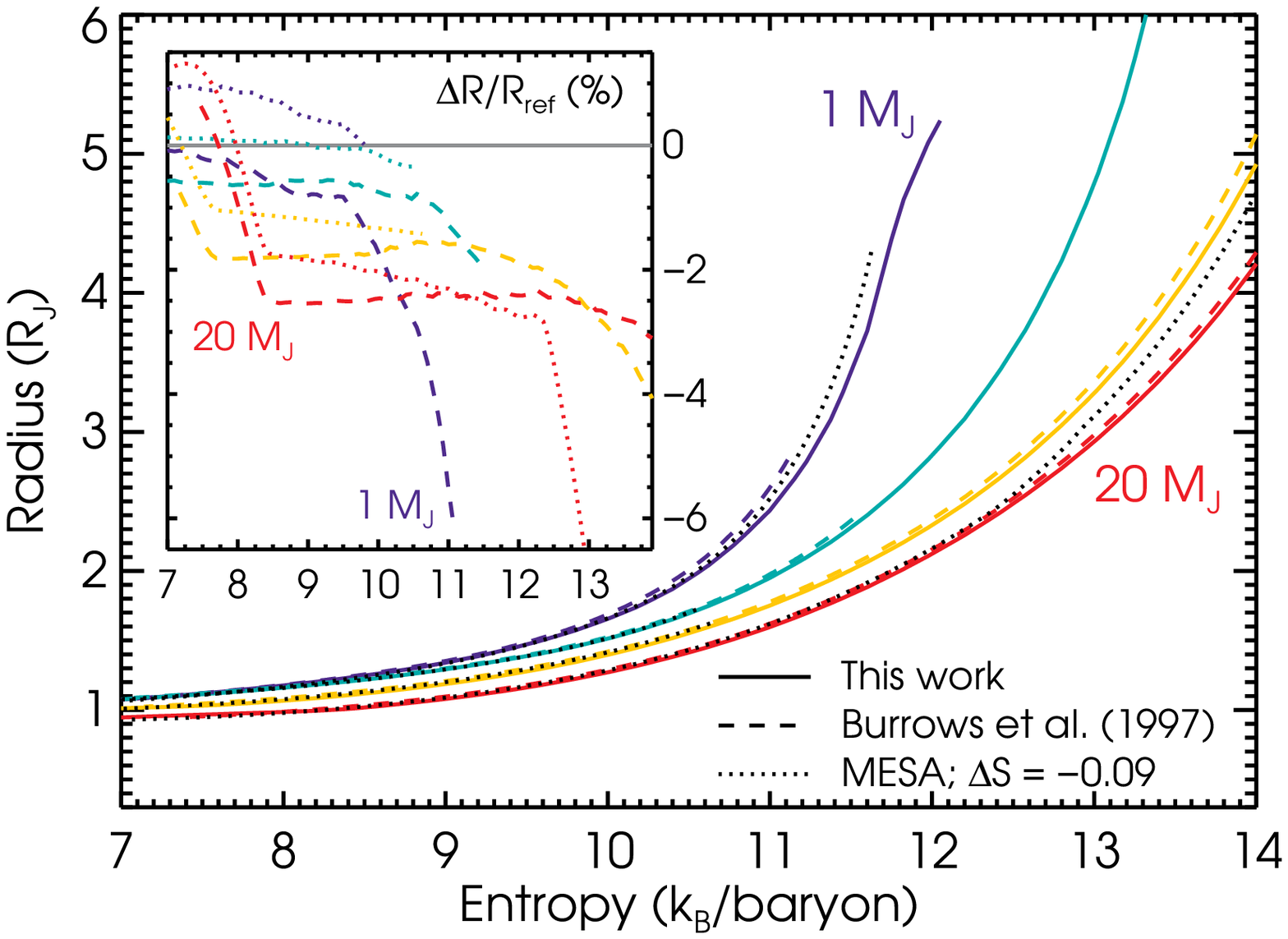}
\caption{
\Ae{
Radii as a function of entropy for $M=1$, 3, 10, and 20~$\MJ$ from top to bottom in our models (solid lines) compared
to those of \citet{burr97} (dashed) %
and \textsc{mesa} (revision~4723; dotted). The entropy values of \protect\citet{burr97}
were shifted by $(1-Y)\ln 2=0.52~\Sunits$ for `thermodynamic reasons' as discussed in Appendix~\ref{sec:Sdiff},
while those of \textsc{mesa} were shifted by the $-0.09~\Sunits$ offset needed to match $L(S)$ (see Fig.~\ref{fig:coolingcurvesSpiegel}).
The inset shows the percentage difference with \protect\citet{burr97} (dashed lines) and \textsc{mesa} (dotted lines)
against entropy. We use for Jupiter's radius $\RJ = 7.15\times10^9$~cm.  %
}
}
\label{fig:SR}
\end{figure}

\Ae{Fig.~\ref{fig:SR} shows $R(S)$ for different masses, using the standard grid with a helium mass fraction $Y=0.25$
and without a solid core. At low entropies, objects of a given mass have a roughly constant radius asymptotically
tending to the zero-temperature value \citep{zapol69,hubb77,ab06},  %
while the radius increases at high entropies. This increase is less pronounced for higher masses,
with more massive objects being smaller at any given entropy.
Finally, there is at a given mass a maximum entropy for which a finite radius is possible,
as the upturn of the curves suggests.
In fact, each curve turns over on to a hot branch (not calculated) where the ion thermal pressure dominates,
leading to distended objects; see for instance the analytic one-zone model of \citet{deloyebild03}.  %
}

\Ae{
The inset of Fig.~\ref{fig:SR} shows the relative difference $\Delta R/R_{\rm ref} = (R-R_{\rm ref})/R_{\rm ref}$
between our radius at a given mass and entropy and that from more detailed calculations,
either \citet{burr97} or \textsc{mesa} (revision~4723; \citealp{paxton11,paxton13}),
over the range where data are available.
There are larger deviation at higher entropies but the overall agreement for masses between 1 and 20~$\MJ$   %
is excellent, with our models systematically a few per cent smaller. %
}

\begin{figure}
\includegraphics[width=84mm]{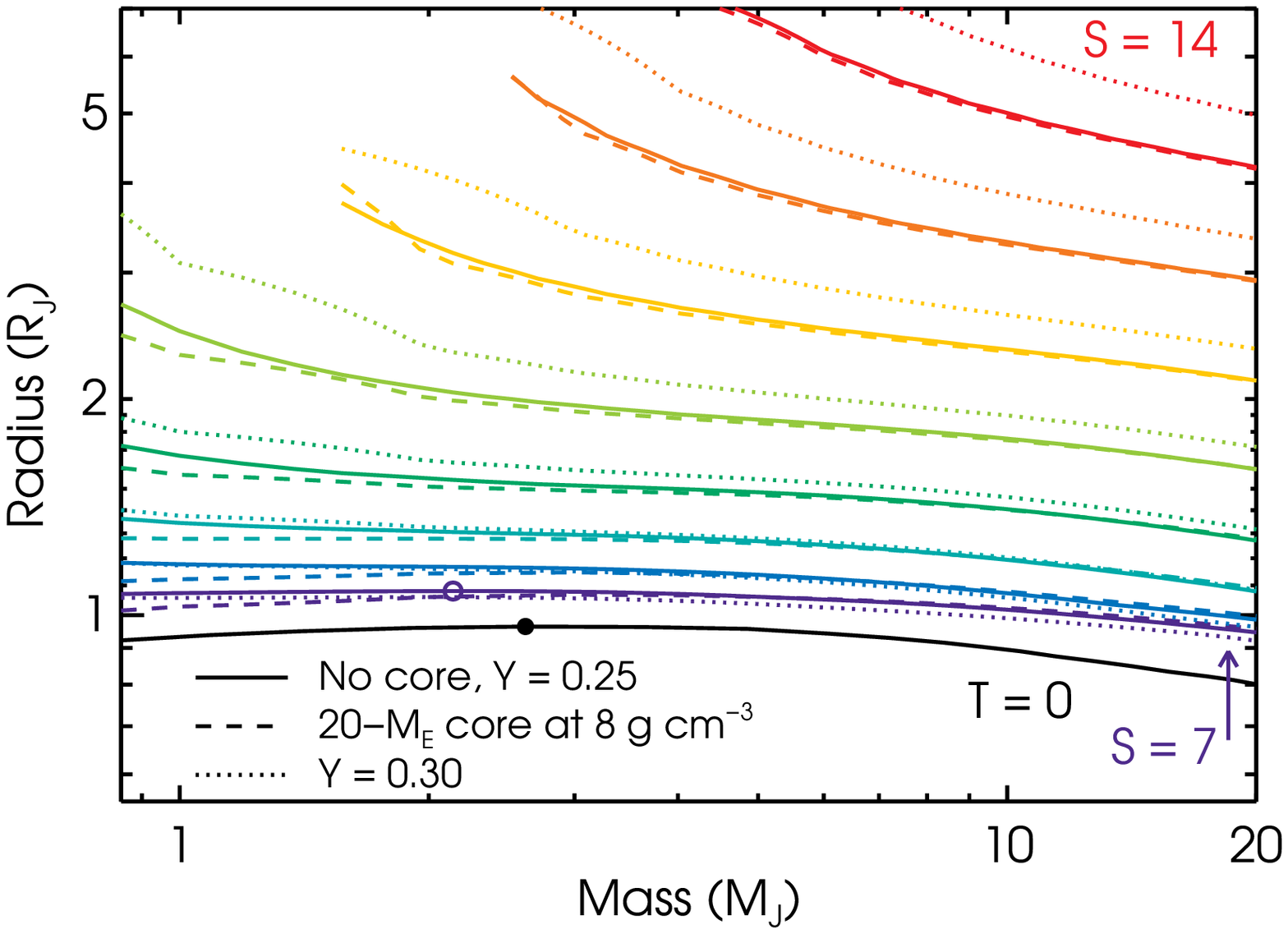}
\caption{
\Ae{
Radii as a function of mass for $S=7$--14 in steps of 1~$\Sunits$ from top to bottom in
the standard grid (without a core and with a helium mass fraction $Y=0.25$; solid lines).
Note the logarithmic vertical scale.
The lowest curve is the $T=0$ result of \citet{zapol69} for $Y=0.25$,
with the peak of $R_{\rm max} = 0.96~\RJ$ at 2.6~$\RJ$ indicated by a filled circle
(note that their table~1 indicates instead 0.98~$\RJ$ at 2.77~$\MJ$).
A ring symbol indicates the peak for $S=7$ in the standard grid.
For comparison, radii in grids with a 20-$\ME$ core and $Y=0.25$ (dashed)
or without a core but with $Y=0.30$ (dotted) are also shown.
}
}
\label{fig:MR}
\end{figure}

\Ae{
In Fig.~\ref{fig:MR} are displayed radii as a function of mass for entropies from 14 down to 7~$\Sunits$.
Objects with $S<10$ have $R<2~\RJ$ for all masses, with radii rapidly increasing at higher entropies. %
Low-entropy planets approach the zero-temperature limit \citep{zapol69},   %
with a maximum\footnote{\Ae{The presence of a turnover in the mass--radius relationship is due to the balance
between the attractive Coulomb forces in the ion-electron plasma and the repulsive Fermi forces between the degenerate electrons.}}
radius of 1.08~$\RJ$ at $2.1~\MJ$ for $S=7$
(cf.\ the $T=0$ result of $R_{\rm max} = 0.98~\RJ$ at 2.77~$\MJ$). %
At higher entropies, there are no solutions below a certain mass, preventing the existence of a finite maximum to $R(M)$
(cf.\ \citealp{deloyebild03}).
}

\Ae{
Fig.~\ref{fig:MR} also shows $R(M)$ curves at fixed $S$ in a grid with a 20-$\ME$ core at a constant density of 8~g\,cm$^{-3}$
and in a grid with a helium mass fraction $Y=0.30$ (and no core). Differences are small when including a solid core
but larger when varying the helium fraction, with differences of the order of tens of per cent
at smaller entropies. %
Note that a realistic equation of state for the core, such as a half-half rock--ice mixture from ANEOS \citep{marl07}
as formation by core accretion may produce,
would yield average core densities closer to $\simeq10$--100~g\,cm$^{-3}$. The effect on the radius should however  %
still be within tens of per cent.
}

\section{A systematic entropy offset in different versions of the SCvH EOS}
\label{sec:Sdiff}

\Ae{Investigating the entropy offset between the models of \citet{spiegel12} and our work (see Section~\ref{sec:modelcomp}),
we noticed that there is a nearly constant entropy offset of 0.51--0.52~$\Sunits$
over a large portion of the $\varrho$--$T$ plane between the entropy of \citet{burr97}\footnote{From the data available at
\url{http://www.astro.princeton.edu/~burrows/dat-html/data/}.}
and that of our \citetalias{scvh} data\footnote{As available at \url{http://aas.org/archives/cdrom/volume5/doc/files5.htm}.}.
As pointed out by D.\ Saumon (2013, priv.\ comm.), this is very probably due to the absence,   %
in some versions of the EOS, of the statistical weight due to the spin states of the proton in the hydrogen partition function.
The term omitted in the entropy (to facilitate comparisons with other equations of state, which usually do not include it; D.\ Saumon 2013, priv.\ comm.)
is of $\kB\ln2$ per proton, which
implies an offset of $(1-Y)\ln2 = 0.52~\Sunits$ for $Y=0.25$. This is almost what we find from the direct comparison, leaving only
a small residual difference $\Delta S\simeq0.07$ from the luminosity comparison ($\Delta S\simeq0.14$ when comparing
with \citealp{marl07}). Thus \citet{burr97}, \citet{spiegel12}, and
\citet{moll12} all use a version which does not include the contribution,
while the published tables include it.}

\Ae{
We emphasize however that an additive constant in the entropy is not important
physically, nor for the evolution of the planet, since only differences in the entropy are meaningful.
The relevance here is only when comparing (initial) entropies as determined
using different models, with the entropy being a convenient label for the adiabat of the interior. In this spirit,
we provide a few values to allow a quick determination of whether a given EOS   %
includes the term or not: for $Y=0.25$ at $P=5.5\times10^{12}$~${\rm erg}\,{\rm cm}^{-3}$ and $T=5\times10^4$~K
(hence $\varrho=1.001$~${\rm erg}\,{\rm cm}^{-3}$), $S=10.557~\Sunits$ according to the interpolated published \citetalias{scvh} tables.
For $Y=0.243$ (0.30), $\varrho = 0.9954$ (1.044)~${\rm erg}\,{\rm cm}^{-3}$ and $S=10.617$ (10.128) at the same $(P,T)$.
To avoid confusion, we encourage publications using the \citet{scvh} equation of state to quote entropy values
as given in the published tables, and to make clear that this is the case.
}

\section{Age, luminosity, and mass constraints}
\label{app}
\subsection{2M1207}
\label{app:2M1207}
\subsubsection{Age and luminosity}
\label{sec:1207_tL}

The first directly-imaged object with a hot-start planetary mass \citep{chauv04,chauv05}
is located 0.8\arcsec\ from its M8 brown-dwarf primary,
a well-studied (\Ae{see \citealp{gizis02} for the report of the discovery} and \citealp{skemer11} for a summary)
member of the young ($8^{+4}_{-3}$~Myr) TW Hydr\ae\ association \citep{chauv04,song06}
at a distance of $53\pm1$~pc (as averaged by \citealp{skemer11} from \citealp{gizis02,gizis07,biller07,mamaj05,mamaj07,ducour08}),
implying a projected orbital separation of $41\pm1$~AU.  %

Determining the luminosity of 2M1207~b is not straightforward.
Photometry is available in the $J$ \citet{mohanty07}, $H$, $K_s$, and $L'$ bands from
NaCo observations at the VLT \citep{chauv04,mohanty07},
at 0.8--1.0, 1.0--1.2, 1.35--1.55, and 1.4--1.8~$\umu$m (similar to $H$) from the {\it Hubble Space Telescope} \citep{song06},
and in the {\it Herschel} SPIRE bands of 250 and 350~$\umu$m \citet{riaz12a,riaz12b}.
\citet{skemer11} also obtained an upper limit at 8.3--9.1~$\umu$m.
Spectroscopy is available at 1.1--1.35~$\umu$m, 1.4--1.8~$\umu$m \citep{chauv04,mohanty07,patience10},
and 1.95--2.5~$\umu$m \citep{patience10} and 2.0--2.4~$\umu$m \citep{mohanty07}, i.e.\ in $JHK_s$ and slightly redwards.
Summing up the fluxes in $J$, $H$, $K_s$, and $L'$ listed in \citet{mohanty07} and using the known distance
gives a luminosity of $2.1\times10^{28}$~\mbox{erg\,s$^{-1}$}, %
or 27--30 per cent of a bolometric luminosity $\log \Lbol/\LSun = -4.68$ to $-4.74$ (see below).  %
A `direct luminosity determination' is thus not possible. %

As summarised by \citet{barman11_1207}, there are inconsistencies between the luminosity of 2M1207~b
and its spectral type, determined by \citet{chauv04} to be L5--L9.5.
With the corresponding bolometric correction BC$_K$ of \citet{goli04},
the $K_s$-band magnitude implies a luminosity of $\log L/\LSun=-4.7\pm0.1$ \citep{barman11_1207}.
The hot-start, equilibrium models of \citet{bara03} then yield from the age and luminosity
an effective temperature of $1010\pm80$~K \citep{barman11_1207}
in disagreement with
$\Teff\simeq1600$~K implied by the spectral type \citep{mohanty07,patience10,patience12}.
However, this second value is questionable. Indeed, the high effective temperature and low luminosity would require an improbably small
radius of $\simeq0.6~\RJ$.

There are two distinct approaches to the solution.
\citet{mohanty07} argue that the actual luminosity has been underestimated due to grey extinction  %
by an almost edge-on disc (the $\simeq 25$-$\MJ$ brown-dwarf primary is accreting; e.g.\ \citealp*{stelz07}).
The correspondingly higher luminosity is reported by \citet{ducour08} as $\log \Lbol/\LSun = -3.8\pm0.1$.
In contrast, \citet{skemer11} argue against the disc explanation and suggest that
thicker clouds than what might be na\"ively expected\footnote{See comment in section~1.2.2 of \citet{marl12}.}
are needed.
\citet{barman11_8799b,barman11_1207} go further and explicitly claim that the problem is with the derived $\Teff$.
They show that an atmosphere model with $\Teff=1000$~K can fit very well
the photometry and spectroscopy if clouds of typical thickness and also, crucially, non-equilibrium
chemisty are included. The latter leads to a heavily reduced methane abundance (by ca.\ two orders of magnitude
at photospheric depths) compared to the chemical-equilibrium clouds and thus to redder colours than expected.
From their best-fitting models, \citet{barman11_1207} estimate a luminosity of $\log L/\LSun =-4.68\pm0.05$,
in agreement with the luminosity derived from BC$_K$.
As \citet{luhm12} notes, differing bolometric corrections for old (field) and young brown dwarfs  %
are thus not the sole explanation.

We adopt the luminosity of \citet{barman11_1207}, in agreement with   %
\citet{neuh12} who report $\log L/\LSun = -4.74\pm0.06$ (with however $\Teff = 1590\pm280$~K).
Using the age of 5--12~Myr and the \citet{bara03} cooling tracks, this luminosity yields
a hot-start mass of 2--5~$\MJ$ as \citet{barman11_1207} state.
\citet{skemer11} give a slightly higher hot-start mass between 5 and 7~$\MJ$ based on $\Teff=1000$~K  %
and the \citet{burr97} models. %
We note that these errorbars match the estimate $\Delta M/M\simeq \frac{1}{2}\Delta t/t\simeq 0.5$
from Section~\ref{sec:MSishape}.

\subsubsection{Mass information}

Mass information of dynamical origin for this two-body system is not available.
Indeed, since the separation implies a period of at least (depending on the eccentricity) 1700~yr,
detectable orbital evolution or change of the velocity amplitude are not expected in the near future \citep{mamaj05},
precluding both astrometry and radial-velocity measurements.

However, the surface gravity is somewhat constrained, which can be used to set an approximate mass upper limit.
\citet{barman11_1207} state that their best-fitting model has $g=10^4$~\mbox{cm\,s$^{-2}$} but do not provide
any sense of how large the uncertainty on this value might be. However, typical errorbars (as in their similar analysis
for HR~8799~b; \citealp{barman11_8799b}) are at least of 0.5~dex.
\citet{mohanty07} found that the fit to both the photometric and spectroscopic data is
rather insensitive to $\log g$ within 3.5--4.5~\mbox{(cm\,s$^{-2}$)},  %
in agreement with the indications of low gravity from the triangular $H$-band spectral shape
and relatively weak Na~{\sc i} absorption (e.g.\ \citealp{allers07,mohanty07}). %
Also, \citet{patience12} fit $J$, $H$, and $K$ spectra with five grids of atmosphere models,
including BT-{\it Settl} \citep*{allard11}, {\it Drift}-{\sc phoenix} \citep{helling08},
and those of Marley et al.\ \citep{ackerman01}.
The Marley et al.\ models yielded a gravity on the edge of their grid
($g=10^{5.0}$~\mbox{cm\,s$^{-2}$}), but the others gave $\log g \simeq 3.5$, 3.5, 4.3, and 5.0~\mbox{(cm\,s$^{-2}$)},
respectively. (With the best-fitting $\Teff=1500$--1650~K, the implied radii are of 0.4--0.7~$\RJ$, well below any theoretical cooling track.)
Even though systematic issues with atmosphere models of young, low-mass brown dwarfs are expected,
we will take these results to suggest tentatively that 2M~1207~b has a low gravity.
As discussed below, the initial entropy on the cold-start branch is $\Si=9.2$, which is thus an {\em upper} limit
to the current entropy. With $S<9.2$ and a reasonable upper limit $\log g<4.0$~\mbox{(cm\,s$^{-2}$)},
the upper bound on the mass is $\simeq7~\MJ$, and for $\log g = 4.35$~\mbox{(cm\,s$^{-2}$)} it is $12.7~\MJ$.
Therefore, we shall assume that the estimates of the surface gravity imply a mass below the deuterium-burning limit,
near 13~$\MJ$ (\citealp{spiegburrmils11,moll12,boden13}; Marleau \& Cumming, in prep.).  %

\subsection{HR 8799}
\label{app:8799}

\subsubsection{Age and luminosities}
\label{sec:8799Lt}

Several properties of HR~8799 let its age be estimated: variability from non-radial oscillations,
low abundance of iron-peak elements, and far-IR excess due to circumstellar dust \citep{marois08}.
Along with its Galactic space motion and position in a Hertzsprung--Russell diagram, these lead
\citet{marois08} to estimate an age range of 30--160~Myr
with a preferred value of 60~Myr, consistent with the 20--150~Myr range of \citet{moor06} based on membership in the Local Association.
Recently, \citet{baines12} used interferometric measurements of HR~8799's radius to derive a stellar mass and age.
They found best-fitting ages of 33 or 90~Myr, depending on whether the star is approaching or moving away from the main sequence.
However, the statistical errorbars, which do not take uncertainties in the stellar models into account,
are considerable in the second case (the 1-$\sigma$ ranges are 20--40 and 40--471~Myr, respectively).
Nevertheless, if \citeauthor{baines12}'s measurement of the stellar radius and the deduced metallicity are correct,
the age range of 1.1--1.6~Gyr from the asteroseismological analysis of \citet{moya10a} would be compromised,
as \citet{baines12} point out.
Indeed, they estimate a near-solar metallicity, which contrasts with the $[{\rm M/H}]\simeq-0.3$ or $-0.1$ result of
\citet{moya10b}, while metallicity is an important input of asteroseismological analyses.
Moreover, there is the statistical argument put forth by \citet{marois08} that massive discs (such as HR~8799's of
0.1~$M_{\oplus}$; \citealp{su09}) are unlikely to be found around older stars.
In our analysis, we shall therefore ignore the 1.1~Gyr result and instead use 20 and 160~Myr
as lower and upper limits, which brackets the ranges reviewed in \citet{moya10a} and \citet{baines12}.

\citet{marois08,marois10} estimated luminosities of $\log \Lbol/\LSun=-5.1 \pm 0.1$ (HR~8799~b),
$-4.7 \pm 0.1$ (cd) and $-4.7 \pm 0.2$ (e) from the known distance of $39.4\pm0.1$~pc and
six infrared magnitudes, covering $\simeq40$ per cent of the bolometric luminosity,
and also from bolometric corrections for brown dwarfs.
For their part, \citet{marl12} recently derived luminosities of
$\log \Lbol/\LSun=-4.95\pm0.06$, $-4.90\pm0.10$, and $-4.80\pm0.09$ for planets b, c\footnote{
In fact, the errorbars on the luminosity of HR~8799~c are clearly non-Gau{\ss}ian, but this will not be taken
into account out of simplicity.}, and d, respectively,
by self-consistently obtaining the radius from evolutionary models.
This contrasts with the usual procedure of optimising $(R/d)^2$ along with $\Teff$ and $\log g$ to fit the photometry,
which yields unphysically small radii of $\simeq 0.8~\RJ$ \citep{barman11_8799b,marl12}.
\Ae{Similarly, in a recent study\footnote{Note also their careful and detailed review of atmospheric modelling efforts for HR~8799~b
and their problems.}
using ``atmospheric retrieval'' (non-parametric determination of the $P$--$T$ and composition structure)
including a simple cloud model, \citet*{lee13} also find for HR~8799~b
a small radius of $0.66^{+0.07}_{-0.04}~\RJ$, which implies with their $\Teff=900_{-90}^{+30}$~K and $\log g=5.0^{+0.1}_{-0.2}$~\mbox{(cm\,s$^{-2}$)}
a bolometric luminosity $\log L/\LSun = -5.57$.
}
Since \citeauthor{marl12} considered only hot starts, i.e.\ fixed $\Si ={\rm high}$,
the cooling tracks are $\Lbol(M,t)$ and $R(M,t)$ relations, which let $R$ and $t$ be uniquely determined from $\log g$ and $\Teff$.
This gives an age of 360~Myr for HR~8799~b and age ranges of 40--100~Myr and 30--100~Myr for c and d, consistent with other literature estimates.
Given the difficulties in obtaining a reasonable fit, \citet{marl12} warn that the first result should not be taken seriously,
and note that the tension in the age would be reduced by considering colder initial conditions.
In our analysis, the luminosity values of \citet{marois08,marois10} will be used since they are standard
and almost or marginally consistent with those of \citet{marl12}.

Finally, an age of 30 or 60~Myr for the system leads \citet{marois10} to derive from the luminosity
and the cooling tracks of \citet{bara03} masses of $(5,7,7,7)$ or $(7,10,10,10)~\MJ$.
The uncertainty in the age implies (see Section~\ref{sec:genconstr}) $\Delta M \simeq 1$--1.3~$\MJ$ for the hot-start values.

\subsubsection{Dynamical stability}

Since it is to date the only directly-imaged multiple-planet system,
HR~8799 has received a considerable amount of attention with regard to its dynamical stability
(e.g.\ \citealp{marois08,reide09,fabry10,moro10,marois10,bergfors11,curr11,sudol12,espo13,curr12_8799};
see reviews in \citealp{sudol12} \Ae{and \citealp{gozmiga13}}).
However, only the most recent studies were able to consider all four planets.
Crucial questions include whether there are two- or three-planet mean-motion resonances
(MMRs; as \citealp{goz09} and \citealp{fabry10} suggest),
what the inclination and eccentricities of the orbits are (for instance, \citealp{lafren09} estimated 13--$23\degr$
for the inclination of HR~8799~b with respect to the plane of the sky) and whether they are co-planar (against which
\citealp{curr12_8799} recently provided evidence, \Ae{while \citet{kenn13} argues in favour}), %
and, naturally, what the masses (including that of the star) are and how long the system should be required to survive.
Solutions are very sensitive to these parameters and even to the numerical integrator used,
as \citet{espo13} note.
The parameter space's high dimensionality makes a proper exploration~-- i.e.\ without artificially-imposed
restrictions as all authors had to assume~--, computationally prohibitive, and trying to include information
about the disc would only make matters worse.

Stability is estimated by using astrometric constraints and numerically evolving the system over time,  %
requiring that it be stable (without collisions nor ejections) for a period equal to its age. 
However, \citet{goz09} and \citet{fabry10} point out that if it is young
with respect to its main-sequence lifetime,    %
HR~8799 could indeed be a transient system undergoing dynamical relaxation.
Therefore, it may not be possible to draw firm conclusions even from the results of a complete analysis.

Nevertheless, if the direction in which these studies point is correct, the planets should have as low masses as allowed,
with however \Ae{somewhat} higher masses permitted if some orbits are resonant.
For this reason, we shall consider as approximate upper limits from stability analyses\footnote{
\Ae{Note that in a very recent study, \citet{gozmiga13} find, using a novel approach which assumes multiple MMRs
but yields masses independently, that broad mass ranges%
which include the hot-start values (though not perfectly for HR~8799~c and~d) are possible: 4--8, 8--12, 8--12, and 7--10.5~$\MJ$ (bcde).
Using these values instead would barely change our derived minimum bounds on $\Si$, lowering some by $\simeq0.1~\Sunits$.}  %
}
masses of 7, 10, 10, and 10~$\MJ$ (bcde).   %
In particular, \Ae{as \citet{gozmiga13} also find}, it seems very likely that none is a deuterium-burning object.

\subsection[beta Pic]{\ensuremath{\bmath{\beta}}~Pic}
\label{app:betaPic}

\subsubsection{Age and luminosity}

The namesake A5 dwarf of the nearby (9--73~pc; \citealp{malo12}) $\beta$~Pictoris moving group has an 
age of $12^{+8}_{-4}$~Myr \citep{zucker01} and asymmetric outer and warped inner 
discs, which have been observed for more than two decades %
\citep[see review in \citealp{lagr11}]{lagr09,lagr12_debris}.
A companion was first detected in $L'$ \citep{lagr09} and subsequently confirmed
at 4~$\umu$m \citep{quanz10} and in $K_s$ \citep{bonn11}.
Very recently, \citet{bonn13} added to these observations photometry in $J$, $H$, and $M'$.
The distance of $19.44 \pm 0.05$~pc\footnote{This is the value obtained
from a re-reduction of {\it Hipparcos} data by \citet{vanL07}. However,
a number of recent studies still use the value of $19.3\pm0.2$~pc \citep{crifo97}.}
to $\beta$~Pic~b implies an orbital separation of 8--9~AU \citep{chauv12,bonn13},
which is the smallest of all low-mass directly-detected objects. An object at this position had been
predicted from the disc morphology by \citet*{freistetter07}.

Until recently \Ae{\citep{bonn13,curr13}}, the only bolometric luminosity estimate for $\beta$~Pic~b was due
to \citet{neuh12}, who report $\log \Lbol/\LSun = -3.90^{+0.07}_{-0.40}$.
They firstly derived, from the $\Teff=1700\pm300$~K of \citet{bonn11}, a spectral type
${\rm SpT} \simeq\textrm{L}2$--T4 using the $\Teff$--SpT relation of \citet[hereafter \citetalias{goli04}]{goli04}.   %
They then estimated from their SpT--BC$_K$ curve   %
a bolometric correction BC$_K=3.3^{+0.15}_{-1.00}$  (T.\ Schmidt 2012, priv.\ comm.).
Thus, the large, asymmetric lower errorbar on the luminosity comes from the large, asymmetric lower
errorbar on the bolometric correction,
which itself is due to the flat $\Teff$--SpT relation between L7 and T4 in \citetalias{goli04}.
A more direct approach to the bolometric luminosity consists of converting the colour to a spectral type
and obtaining from this a bolometric correction. Also using the fits\footnote{\Ae{It was brought to our attention
that there is an extension of \citetalias{goli04} by \citet{liu10BC},
who use updated spectral types and removed binary systems from the sample.
However, differences in BC$_K$ only begin appearing later
than $\simeq{\rm L}1$, whereas the SpT we consider for $\beta$~Pic~b is L1--T0 (\citealp{bonn11};
cf.\ the constraints of L0--L4 by \citealp{bonn13} or L2--L5 by \citealp{curr13}).
According to the fit of \citet{liu10BC}, ${\rm BC}_K({\rm L})=3.05$~mag,
while \citetalias{goli04} gives 2.99~mag; the difference (0.06~mag) is less than the root-mean-square fit residuals
(0.08~mag and 0.13~mag, respectively).}} of \citetalias{goli04},
this gives the same luminosity as found by \citet{neuh12} but with a smaller
lower errorbar of $0.12$~dex. One should however note %
that ${\rm BC}_K ({\rm SpT})$ is not a monotonic function \citepalias[see fig.~6a of][]{goli04},
so that the errorbars are strongly non-Gau{\ss}ian. With a maximum ${\rm BC}_K$ of 3.3~mag near \Ae{L3.5},
$\log \Lbol/\LSun$ cannot formally be above $-3.9$~dex.
However, the BC$_K$--SpT relation of \citetalias{goli04} was derived for field dwarfs \citep[see also][]{steph09},
and the spectral classification of young objects is not yet well understood nor, in fact, well defined
\citep[see e.g.][]{liu11,faherty12}.
Keeping in mind these uncertainties in interpreting the photometry,
we shall use for the analysis $\log \Lbol/\LSun = -3.90^{+0.05}_{-0.12}$,
where the upper errorbar reflects the residuals of the \citetalias{goli04} fit.
\Ae{This is at some variance with the value of $\log L/\LSun = -3.80\pm0.02$ of \citet{curr13}
but compares favourably with $\log L/\LSun = -3.87\pm0.08$ from \citet{bonn13}.}

\subsubsection{Mass information}

The object $\beta$~Pic~b is particularly interesting because it is the first directly-imaged companion for which radial-velocity data
are also available \citep{lagr12_rv}.
Using new and archival data spanning eight years and thanks to the high inclination of the system ($88\pm2\degr$; \citealp{chauv12}),
\citet{lagr12_rv} were able to place tentative lower mass limits of 1--2$~\MJ$, which is fully consistent with all
reasonable age and luminosity combinations, even allowing for very large lower errorbars on the latter.
However, \citeauthor{lagr12_rv}'s upper limit of 10--$25~\MJ$, with $12~\MJ$ for the most probable orbit of 9~AU %
\citep{lagr09,chauv12} is an important result
which excludes high-mass solutions and puts the object quite likely in the planetary (non-deuterium burning) range.

\Ae{We conclude with a brief digression.} %
To \Ae{the} rarity \Ae{of objects observable simultaneously in radial velocity (RV) and in direct imaging} contribute both intrinsic detection biases
-- direct imaging favours planets further out from their star, resulting in a small RV signal -- as well as selection biases
-- target stars are usually chosen based on the presence of a disc, which implies that the systems are preferentially seen face-on.
A further hindrance is that young stars -- young systems being of \Ae{greater interest because of a smaller brightness contrast} -- are usually active
and thus less amenable to radial-velocity measurements. \Ae{See also \citet{lagr13} for a discussion of radial-velocity
searches around young nearby stars and example prospects of coupling to direct imaging.}

\end{document}